\documentclass[a4paper,11pt]{article}
\usepackage{jinstpub}

\usepackage{natbib}
\usepackage{graphicx}
\usepackage{url}
\usepackage{xcolor}
\usepackage{amsmath}
\usepackage{bm}
\usepackage{enumerate}
\usepackage{subcaption}
\usepackage{lineno}
\usepackage{soul}
\usepackage{tikz}
\usepackage{ragged2e}
\def\checkmark{\tikz\fill[scale=0.4](0,.35) -- (.25,0) -- (1,.7) -- (.25,.15) -- cycle;} 
\usepackage{pgfplots}
\pgfplotsset{compat=1.11}
\usepackage{lineno}
\usepackage{floatrow}

\usepackage{tabularx}
\usepackage{xfrac}

\usepackage{fancyhdr}
\usepackage[yyyymmdd,hhmmss]{datetime}
\usepackage{float}
\floatstyle{plaintop}
\restylefloat{table}

\usepackage{sidecap}
\usepackage{siunitx}

\usetikzlibrary{calc,angles,positioning,shapes.geometric,arrows,patterns}

\usepackage{hyperref}
\hypersetup{colorlinks=true, linkcolor=blue, citecolor=blue}

\tikzstyle{Stage_process1} = [rectangle,draw, text width=8em, text centered, rounded corners, minimum height=5em, draw=blue]

\tikzstyle{alto_process} = [rectangle,draw, text width=7em, text centered, rounded corners, minimum height=5em]

\tikzstyle{StageB} = [rectangle, minimum width=3cm, minimum height=1cm, text centered, text width=4cm, draw=black]

\tikzstyle{StageB_decision} = [diamond, minimum width=3cm, minimum height=1cm, text centered,text width=1.8cm, draw=black]

\tikzstyle{arrow} = [thick,->,>=stealth]
\tikzstyle{dash_line} = [thick,dashed,>=stealth, blue]

\newcommand*{\rdash}{--}

\newcommand*{\ack}{Acknowledgements}

\begin{document}

\title{\boldmath{Signal extraction in atmospheric shower arrays \\ designed for $\rm 200\,GeV-50\,TeV$ $\gamma$-ray astronomy}}

\author[a,1]{M.~Senniappan,
\note{Corresponding authors.}}
\author[a,1]{Y.~Becherini,}
\author[b,a]{M.~Punch,}
\author[c]{S.~Thoudam,}
\author[a]{T.~Bylund,}
\author[a]{G.~Kukec~Mezek,}
\author[d]{J.-P.~Ernenwein}

\affiliation[a]{Department of Physics and Electrical Engineering, Linnaeus University, 35195 V\"axj\"o, Sweden}
\affiliation[b]{Astroparticule et Cosmologie (APC), CNRS, Universit\'e Paris 7 Denis Diderot, 10, rue Alice Domon et Leonie Duquet, F-75205 Paris Cedex 13, France}
\affiliation[c]{Department of Physics, Khalifa University, PO Box 127788, Abu Dhabi, United Arab Emirates}
\affiliation[d]{Aix Marseille Univ, CNRS/IN2P3, CPPM, Marseille, France} 

\emailAdd{mohanraj.senniappan@lnu.se}
\emailAdd{yvonne.becherini@lnu.se}
% Reintroduced the \received and \accepted commands from AASTeX v5.2
%\received{February 20, 2021}
%\revised{}
%\accepted{}

\abstract{
We present the SEMLA (Signal Extraction using Machine Learning for ALTO) analysis method, developed for the detection of $\rm E>200\,GeV$ $\gamma$ rays in the context of the ALTO wide-field-of-view atmospheric shower array R\&D project. 
The scientific focus of ALTO is extragalactic $\gamma$-ray astronomy, so primarily the detection of soft-spectrum $\gamma$-ray sources such as Active Galactic Nuclei and Gamma Ray Bursts. 
The current phase of the ALTO R\&D project is the optimization of sensitivity for such sources and includes a number of ideas which are tested and evaluated through the analysis of dedicated Monte Carlo simulations and hardware testing. 
In this context, it is important to clarify how data are analysed and how results are being obtained. 
SEMLA takes advantage of machine learning and comprises four stages: initial event cleaning (stage A), filtering out of poorly reconstructed $\gamma$-ray  events (stage B), followed by $\gamma$-ray signal extraction from proton background events (stage C) and finally reconstructing the energy of the events (stage D). 
The performance achieved through SEMLA is evaluated in terms of the angular, shower core position, and energy resolution, together with the effective detection area, and background suppression. Our methodology can be easily generalized to any experiment, provided that the signal extraction variables for the specific analysis project are considered.
}


\maketitle
\flushbottom

\section{Introduction}

\begin{figure*}[t]
    \centering
    \includegraphics[width=0.43\textwidth] {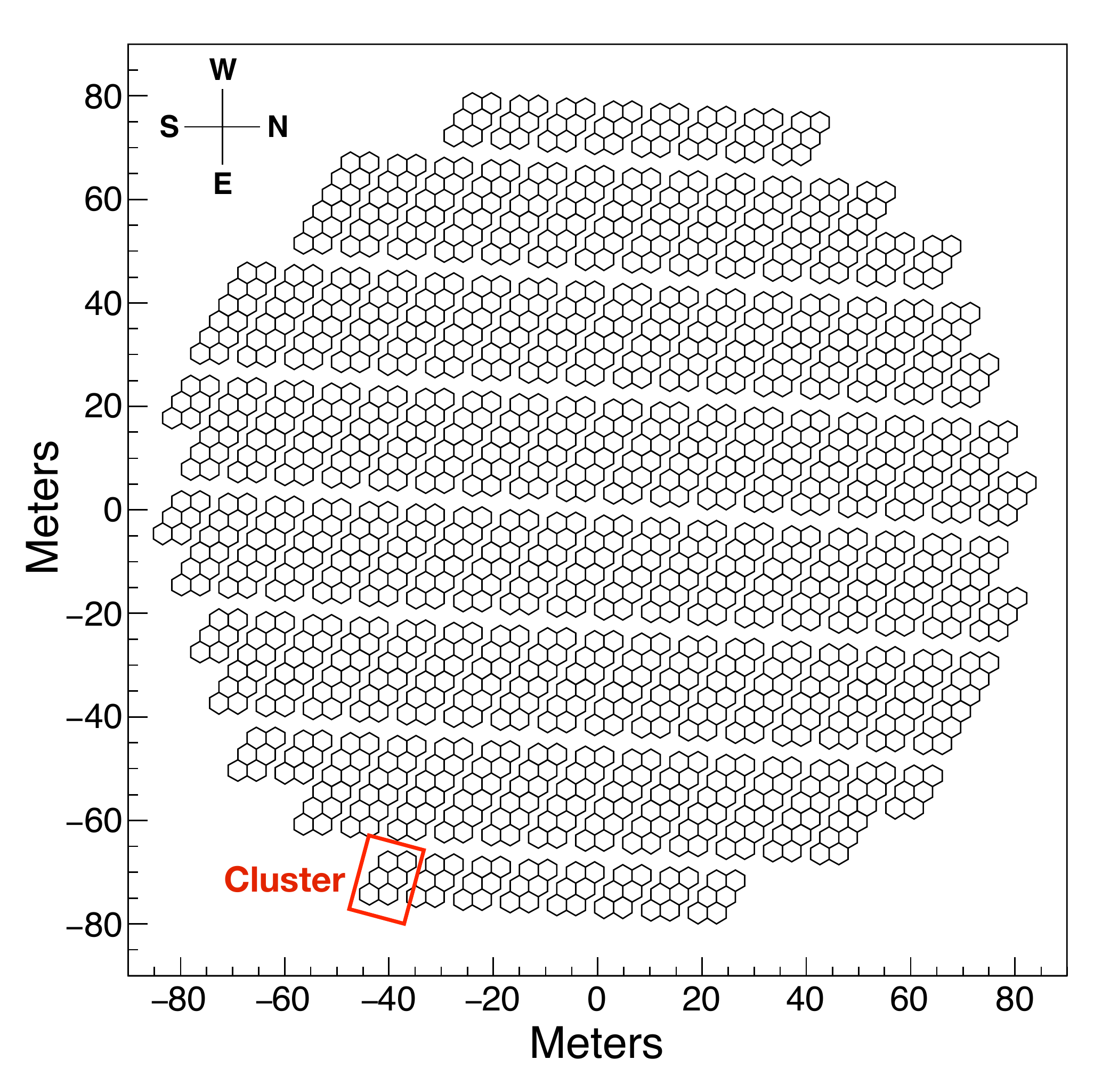}
    \hspace{0.1cm}
    \includegraphics[width=0.55\textwidth] {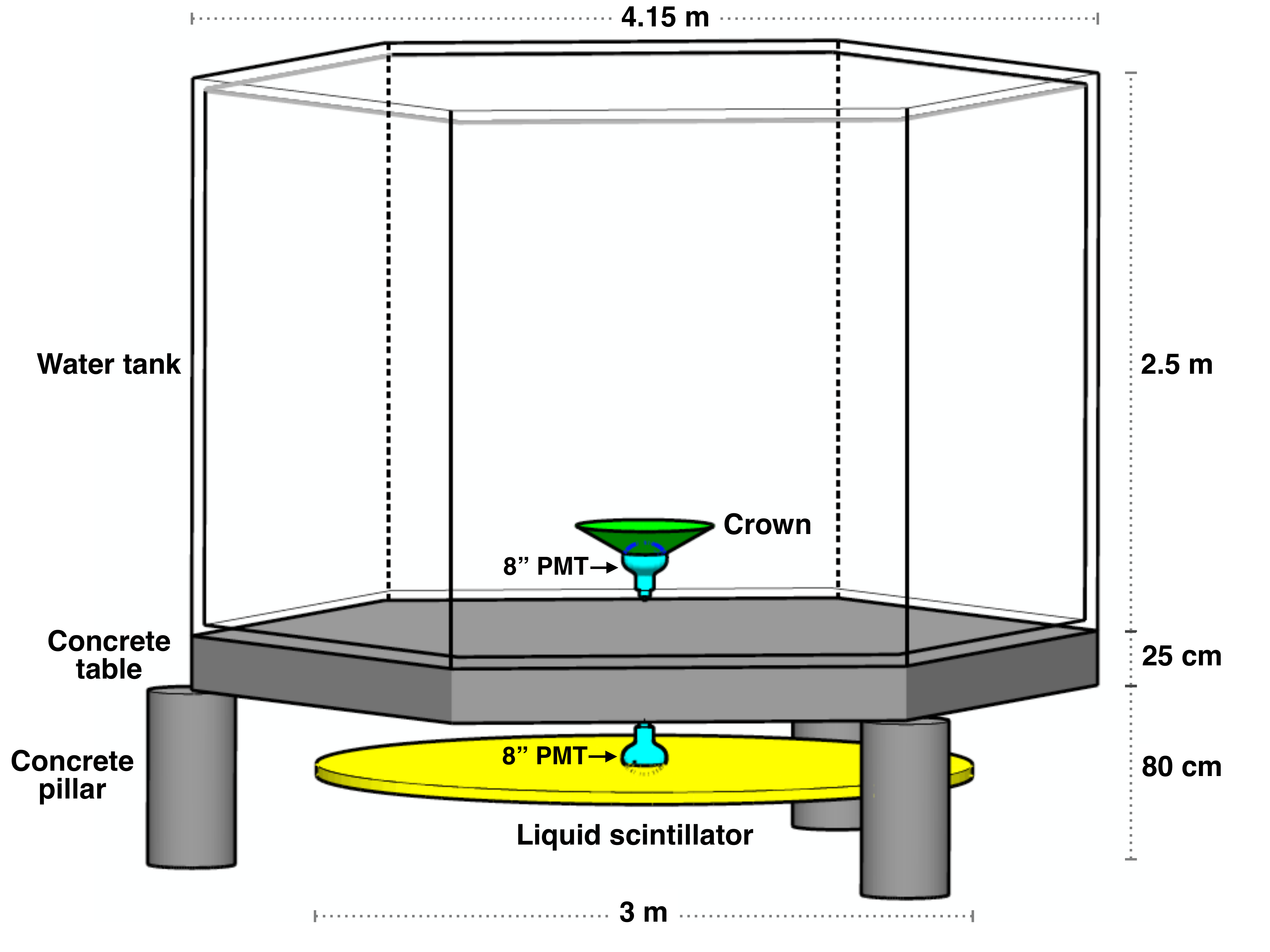}
    
    \caption{\small{\textbf{ALTO design.} \emph{Left panel:} The layout of the proposed ALTO array of $\rm 160\,m$ diameter. A \emph{cluster} consists of six ALTO units as highlighted in the red box. \emph{Right panel:} The geometry of an ALTO unit with its dimensions. A unit consists of a water Cherenkov detector (WCD), a liquid scintillator detector (SD) underneath, separated by a concrete slab. }}
    
    \label{fig:alto-array}
    
\end{figure*}

ALTO is an R\&D project focused on the design of a very-high-energy (VHE, here for $\rm{E>200\,GeV})$ $\gamma$-ray observatory \cite{alto1, alto2}. 
The project belongs to the family of air-shower arrays, in which $\gamma$-rays are reconstructed from the detection of their cascade of secondary particles generated in the atmosphere. 
In more detail, the interaction of a VHE $\gamma$-ray with the electric field of an atmospheric nucleus gives rise to an electromagnetic cascade of relativistic particles which consists of electrons, positrons, and lower-energy $\gamma$-rays. 
In contrast, the interaction of a cosmic-ray with such a nucleus results in charged mesons which decay into muons and neutrinos, and neutral pions which lead to electromagnetic sub-showers. The particles contained in the atmospheric cascades can be detected by ground-based arrays like the one proposed here, which is designed to be composed of 1242 detection units positioned in a circular area (radius $80\,\rm m$), see figure~\ref{fig:alto-array} (left panel). 
Each ALTO unit consists of a hexagonal Water Cherenkov Detector (WCD) positioned on a concrete slab, with a cylindrical liquid Scintillator Detector (SD) underneath, see figure~\ref{fig:alto-array} (right panel). 
The WCDs are primarily used for the detection of  particles in the cascade, while the SDs are conceived for muon tagging, and the concrete layer acts as an absorber to prevent most electrons and positrons from reaching the SDs.

From the data analysis perspective, irrespective of the specific scientific context, every experiment faces two major challenges: the selection of well-reconstructed events and the extraction of \emph{signal} from \emph{background}. 
For more than a decade, multi-variate algorithms have been successfully employed to this end in $\gamma$-ray astronomy\footnote{As, for instance, Boosted Decision Trees (BDT) and Neural Networks (NN).} \cite{new_analysis_strategy, CTA_small_prototype, act_nn, HAWC_NN}. 
In this paper, we present a new analysis strategy also exploiting the advanced methods offered by Machine Learning. 
The analysis strategy implemented in this context is called SEMLA (Signal Extraction using Machine Learning for ALTO) and it comprises several stages, summarized in figure~\ref{fig_flow:ALTO_stages}. 
With SEMLA, we aim to achieve a good performance in terms of angular and energy resolution, energy bias, effective detection area and background suppression for an optimized sensitivity in the energy region of interest (from $\SI{200}{\GeV}$ to $\SI{50}{\TeV}$).

This paper comprises eight sections describing the implementation of the SEMLA analysis method. 
Section \ref{sec:simu_reco} introduces the Monte Carlo simulation and reconstruction technique developed for the ALTO R\&D phase. 
Also in section \ref{sec:simu_reco}, we present the ALTO concept briefly, since a more extensive paper about the project is in preparation and the work presented here focuses on data analysis aspects.
The importance of the analysis stages is presented in section \ref{sec:Analysis}. 
This section also introduces the Hillas parameters used in the analysis and explains initial data preparation. 
Section \ref{sec:data_clean} describes the event cleaning. 
An in-depth description of the classification tasks used in the filtering of poorly reconstructed events and in the $\gamma$/proton separation is presented in sections \ref{sec:select_best} and \ref{sec:gamma-hadron} respectively. 
Section \ref{sec:energy_reco} presents the primary energy reconstruction of $\gamma$-like events using a regression method. 
Finally, we summarise the performance of the SEMLA analysis approach in section \ref{sec:performance} and conclude with a discussion on possible future improvements in section \ref{Outlook}.

\section{Monte Carlo simulation and shower reconstruction method}
\begin{figure*}[t]
\small{
\resizebox{14cm}{5.5cm}{
\begin{tikzpicture}[node distance = 3cm]

  \node (show_sim) [alto_process] {CORSIKA shower simulation};
  \node (det_sim) [alto_process, right of = show_sim] {GEANT4 detector simulation};
  \node (reco) [alto_process, right of = det_sim] {Shower reconstruction};
  \node[color=blue] (ana) [alto_process, right of = reco] {SEMLA analysis};  
  \node (high_ana) [alto_process, right of = ana] {Performance evaluation};
  
  \node (pA) [Stage_process1, below of = show_sim, yshift = -0.5cm] {A. Data cleaning via simple cuts (section \ref{sec:data_clean})};
  \node (pB) [Stage_process1, right of = pA, xshift = 0.8cm] {B. Filtering poorly reconstructed events via \\ classification (section \ref{sec:select_best})};
  \node (pC) [Stage_process1, right of = pB, xshift = 0.8cm] {C. $\gamma$/hadron separation via \\ classification (section \ref{sec:gamma-hadron})};
  \node (pD) [Stage_process1, right of = pC, , xshift = 0.8cm] {D. Energy evaluation via regression (section \ref{sec:energy_reco}) };
  
  \draw [arrow] (show_sim) --  (det_sim);
  \draw [arrow] (det_sim) --  (reco);
  \draw [arrow] (reco) --  (ana);
  \draw [arrow] (ana) --  (high_ana);
  
  \draw [arrow] (pA) --  (pB);
  \draw [arrow] (pB) --  (pC);
  \draw [arrow] (pC) --  (pD);
  
  \draw [dash_line] (ana.-148) --  (pA.149);
  \draw [dash_line] (ana.-33) --  (pD.33);
\end{tikzpicture}
}
}
\caption{\label{fig_flow:ALTO_stages} \small{The workflow from the simulation step to the performance evaluation with the expansion of the different analysis stages in SEMLA.}}

\end{figure*}
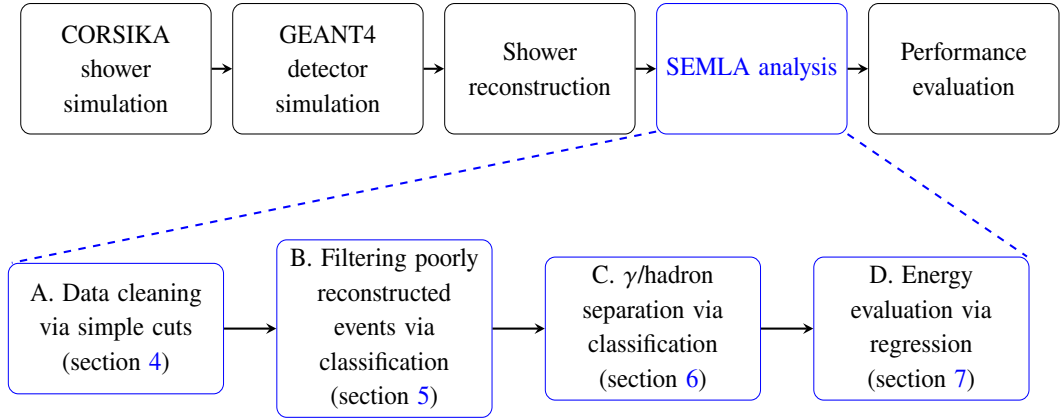

The full workflow used in our study of the performance of the ALTO detector is shown in figure~\ref{fig_flow:ALTO_stages}. 
In this section we will present how the first three steps (atmospheric shower simulation, detector simulation and event reconstruction) are carried out.

\label{sec:simu_reco}

\subsection {Simulations of particle cascades in the atmosphere}
\label{subsec:sim_atmos}

The interaction of primary particles at the top of the atmosphere and the subsequent shower development is simulated using the CORSIKA software (version $7.4387$) \cite{Heck}. 
$\gamma$ rays are simulated at a fixed zenith angle of 18$^{\circ}$ and an azimuthal angle of 0$^{\circ}$ with a differential power law spectrum of ${\rm d}N/{\rm d}E=E^{-2}$, thus representing a simulated ``point source''. To give an idea of the corresponding rates for such a $\gamma$ ray source, in the following estimations we consider a ``Pseudo-Crab'' as a bright, stable source with a spectral index close to that simulated for the $\gamma$ rays.\footnote{This is named ``Pseudo-Crab'' because the Crab is a northern hemisphere source, and there is a choice of spectral shapes which have been fitted by various VHE observatories.  Here, we consider the same power law for the Crab spectrum as in \cite{MAGIC_Crab} but without their exponential cut-off; as parameters we take their differential flux at $\SI{1}{\TeV}$ of $\SI[mode=text]{3.8d-7}{m^{-2}s^{-1}TeV^{-1}}$, and an $\alpha$ index of $-2$ (which is close to their $-2.21$).}

In order to evaluate the \emph{background} of cosmic rays in the same region of the simulated point-like source, we simulate proton interactions in the atmosphere in a wider range of zenith angles, 15--$21^{\circ}$, and azimuthal angles, 0--$360^{\circ}$, with a power law spectrum ${\rm d}N/{\rm d}E=E^{-2.7}$. 
For approximate estimates of the resulting cosmic-ray rates in the following, we re-scale these protons to the all-particle cosmic-ray flux.\footnote{We take the all-particle cosmic-ray flux given by \cite{ST_all_particle}, with differential spectral index at $\SI{1}{\TeV}$ of $\SI[mode=text]{0.21}{m^{-2}{sr}^{-1}s^{-1}TeV^{-1}}$ but taking the spectral index as the proton-simulated $-2.7$ rather than $-2.6$ for their all-particle spectrum.  We consider this to be conservative, as the protons make up $\sim 40\%$ of the all-particle spectrum for this energy range, but we expect that the heavier species will be more easily rejected.}
The total livetime of the simulated \emph{background} then corresponds to $\sim13\,\rm minutes$ of all-particle cosmic ray observations in a realistic case of the data taking by a full 1242-unit array. 

The characteristics of the air shower simulations used in this work are summarized in table~\ref{table:simulation}. 
The secondary particles in the air shower development are simulated for an observation altitude of $\SI{5.1}{\km}$, in the US standard atmosphere, with the horizontal/vertical components of the magnetic field taken as $\SI{21.12}{\micro\tesla}$ and  $-\SI{8.25}{\micro\tesla}$ respectively.

The final choice of the ranges in energy, impact parameter, and zenith angle for protons were adjusted as we proceeded with the analysis. In this process, if any of the simulation ranges are found to be insufficient, they are simulated by extending these characteristics.

The choice was made to simulate $\gamma$ rays above $\SI{10}{\GeV}$.  From the response to the full SEMLA analysis stages shown in figure~\ref{fig:Energy_distributions} (left panel) it is apparent that a negligible number of $\gamma$ rays remain after SEMLA application at the lowest energies.
In fact, less than 2.5 per million of the $\gamma$-rays simulated below $\SI{80}{GeV}$ are found to pass the SEMLA analysis cuts (which gives $\SI{0.4}{\per day}$ for the Pseudo-Crab source), and since these in any case are reconstructed with energies over $\sim\SI{150}{\GeV}$ (see section\ \ref{sec:energy_reco}), they do not change significantly the instrument response.
Therefore, $\SI{10}{\GeV}$ represents an overly-conservative limit for the simulations.  

For protons, we performed the simulations down to $\SI{60}{\GeV}$.  
After implementing the SEMLA analysis cuts (see figure \ref{fig:Energy_distributions}, right panel), there remain 0.3 proton events per million simulated below $\SI{80}{\GeV}$, which 
corresponds to a cosmic-ray background rate of approximately 4 per $\rm deg^2$ per day. So, we conclude that the background from below that energy should be negligible.

Each shower is simulated with a random shower core position, relative to the centre of the array, in a disc of radius $130\,\rm m$ for $\gamma$-rays and $184\,\rm m$ for protons. After applying the pre-cuts described in section \ref{sec:select_best}, almost no events with impact parameters near
the perimeter of those discs are retained.

In more detail, to verify that simulations have been performed up to sufficiently large impact parameters, we checked the distribution of impact parameters for events passing all SEMLA analysis cuts.  For $\gamma$ rays, this drops rapidly after $80\,\rm m$ (the edge of the array), with no events remaining whose shower core was near the perimeter of the simulated area. 
For protons, this drop at $80\,\rm m$ also occurs, but with a slower exponential drop-off to zero before the $184\,\rm m$ limit. Note that we had initially simulated protons with the same range as the $\gamma$-rays, but extended their range to take this slow drop-off into account.

The proton events that pass the SEMLA analysis cuts were found to have their angular origin reconstructed quite well, with the root-mean-square (RMS) of the zenith angle error being $0.7^\circ$.  In the analysis of $\gamma$-ray point sources, to reject the cosmic-ray background, we apply an angular selection cut based on the angular resolution for $\gamma$ rays which is at most $1^\circ$, as shown in figure \ref{fig:StageC_selected_angular_resolution}.  If there were proton events outside the simulated zenith angle range ($18 \pm 3)^\circ$ which would have fallen into the $\gamma$-ray angular selection cut region, then they would increase the background.  However, we find the fraction of protons with a zenith angle reconstruction error $> 2^\circ$ (i.e.\ $3^\circ - 1^\circ$) to be 1.2\%, and we conclude that the zenith angle range simulated for the proton background is sufficient.

\begin{table}[t]
    \centering
    \small
    \caption{\small{\textbf{Simulation details.} CORSIKA shower simulation characteristics and number of events simulated. 
    $ \theta_{\rm T} $  and $ \phi_{\rm T} $ represent the true zenith and azimuthal angle respectively.}}
    \label{table:simulation}
    \begin{tabular}{ccccccc}
        \hline
        \textbf{Primary} & \textbf{Spectral} & \textbf{Energy} & \boldmath{$\theta_{\rm T}$}& \boldmath{$\phi_{\rm T}$} & 
        \textbf{Impact} & \textbf{Events} \\ 
        
        \textbf{type} & \textbf{index} & \textbf{range} & \textbf{[deg]}& \textbf{[deg]} & \textbf{parameter} & \textbf{(x\boldmath{$10^6$)}} \\
        
        & & \textbf{[TeV]} & & & \textbf{[m]} & \\ 
        
        \hline
            $\gamma$ ray & -2 & 0.01\rdash100 & 18 & 0 & 0\rdash130& 34 \\
            proton & -2.7 & 0.06\rdash100 & 15\rdash21 & 0\rdash360 & 0\rdash184 & 224\\
        \hline
    \end{tabular}

    \end{table}
    
\subsection{Simulation of the detector response}
The GEANT4 (version \emph{4.10.02.p02}) libraries \cite{GEANT4} are  used  for detector response simulations of the ALTO array, with 1242 units covering a $160\,\rm m$ diameter area, see figure~\ref{fig:alto-array} (left panel).
The array is further subdivided into \emph{clusters}, each consisting of 6 detection units (or 12 detectors in total), where a single unit, shown in figure~\ref{fig:alto-array} (right panel), is composed of a water tank (WCD) and a scintillator (SD), separated by a concrete slab. 

\textbf{Water Cherenkov detector (WCD)}: Each WCD is painted black inside and is filled with $\rm25\,m^3$ of water. 
A super-bialkali $8^{\prime\prime}$ Hamamatsu Photo Multiplier Tube (PMT), positioned at the bottom of the WCD, records Cherenkov photons generated by relativistic particles in water. 
Having a black interior, the WCD is optimized for the detection of direct Cherenkov photons, which are emitted in water at an angle of $\theta_{C}=42^\circ$. 
A reflective crown helps in increasing the collection of direct photons which would otherwise be lost, recovering up to $\sim~40$\% of photons. 

\textbf{Scintillators (SD)}: The interaction of an incoming particle with organic liquid scintillator (Linear Alkyl Benzene, LAB, plus wavelength shifters), gives rise to the emission of photons whose wavelength matches the PMT photocathode sensitivity.
The SDs contain a standard-bialkali $8^{\prime\prime}$ Hamamatsu PMT.

GEANT4 is used to simulate the interaction of particles inside the detectors and ray-trace the optical photons. 
The output of the simulation are the number of photons and the photon arrival times at the PMT level. They are then processed with our own PMT response simulation to generate the electronic waveforms of the signals.
A cluster is considered to be \emph{triggered} if at least two detectors among twelve (``2/12'') have an amplitude exceeding $\SI{20}{\mV}$. In such a case, the signals of the twelve channels are reduced to their time of maximum (time of the peak of the waveform) and integrated charge ($S_{\rm pe}$), which are stored and considered in the next steps. Signals from the non-triggering clusters are discarded. 
In the ``2/12'' trigger condition, we therefore use both WCD and SD signals. 
We discuss the performance of the chosen trigger settings in section \ref{Outlook}.

In this paper, the important shower parameters that we use to characterise an event are the shower core position $(X,Y)$, the impact parameter $b \; (=\sqrt{X^2+Y^2})$, the arrival direction $(\theta, \phi)$  and the energy of the primary particle $E$. Hereafter, subscripts $T$ or $R$ will be used to represent the true or the  reconstructed variables of the shower parameters.
  
\begin{figure*}[t]
\centering
\begin{small}
\includegraphics[width=0.47\textwidth]{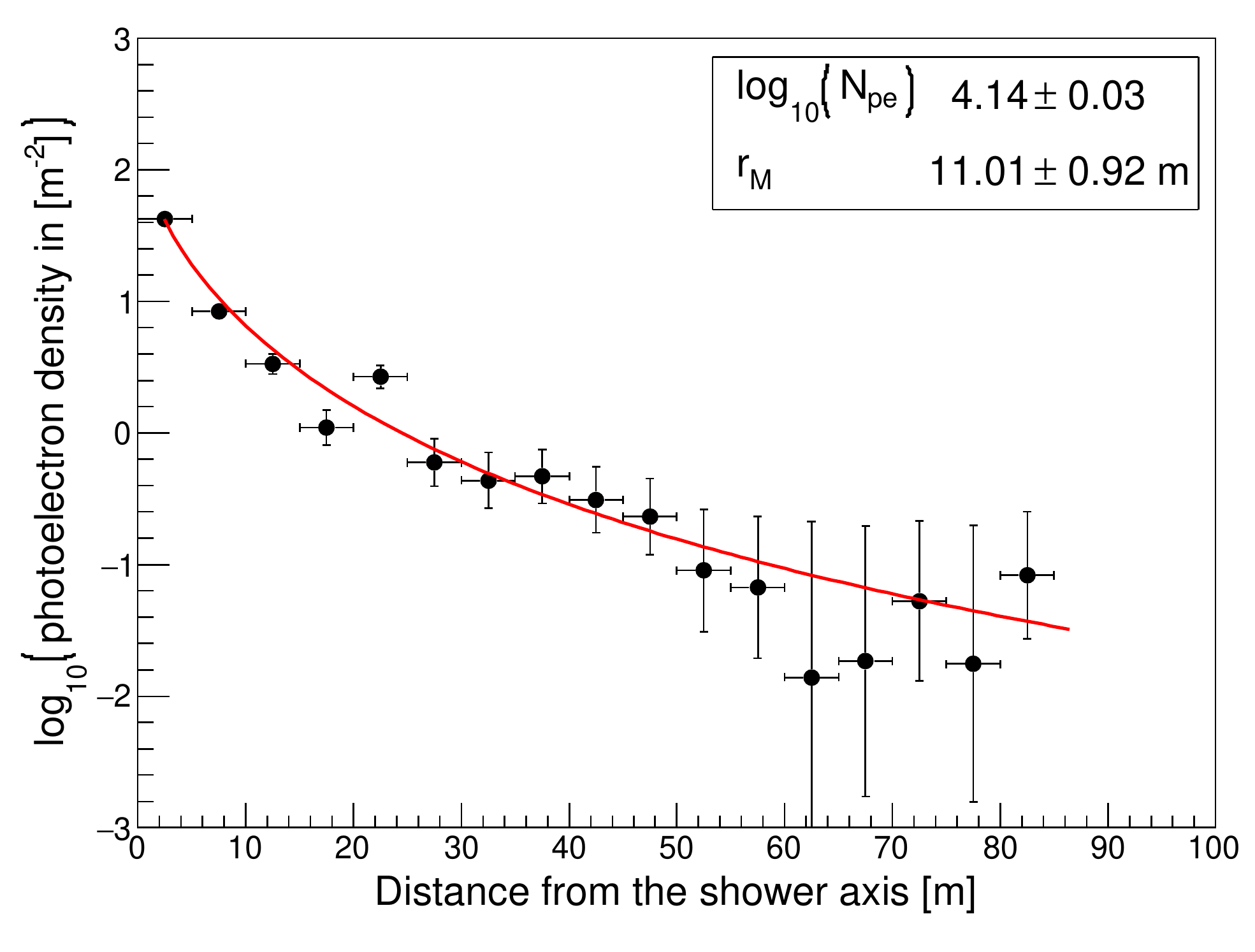}
\hspace{0.5cm}
\includegraphics[width=0.47\textwidth]{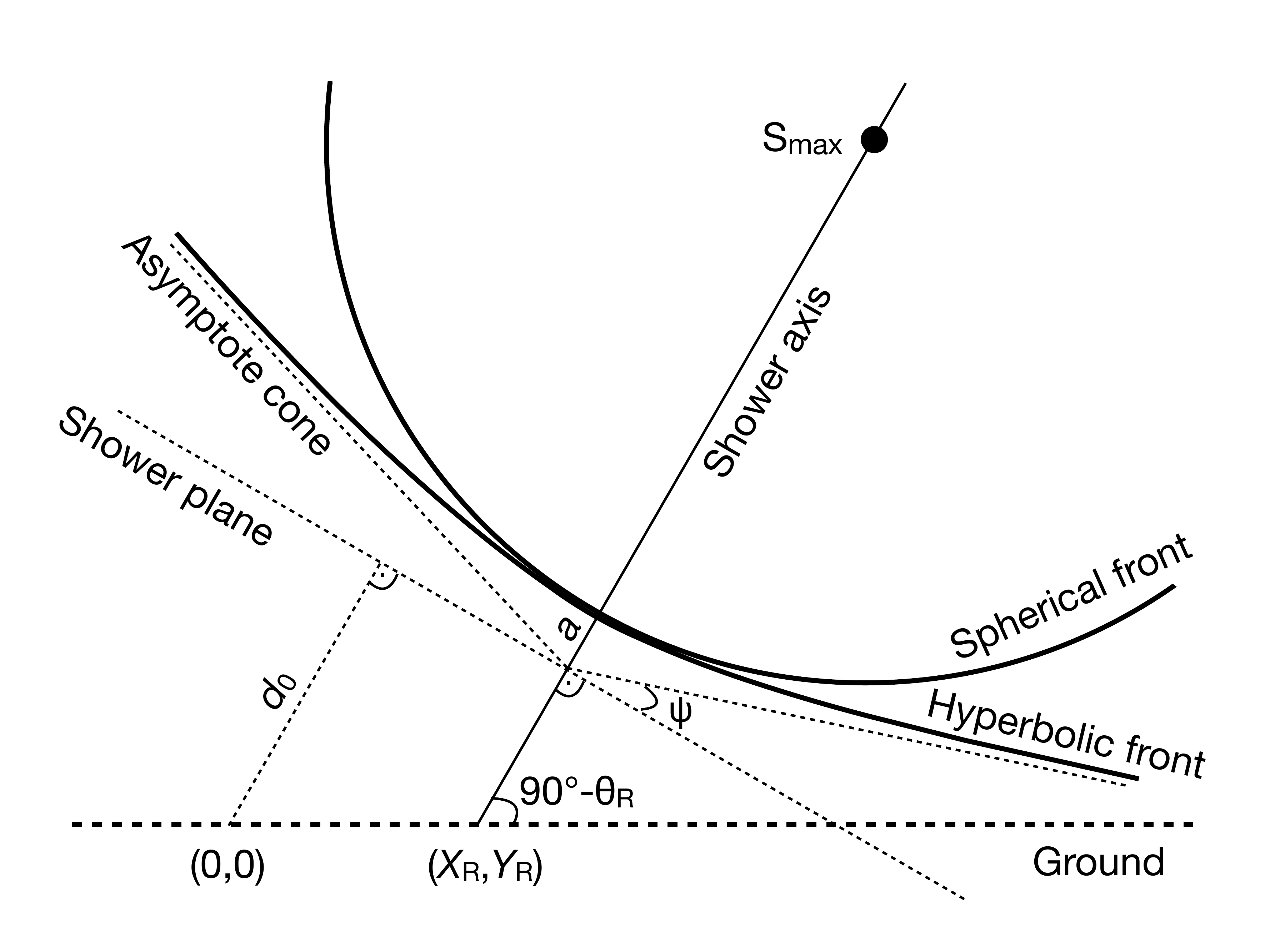}
\caption{\small{\label{fig:wavefront} \emph{Left Panel:} Lateral distribution of a $\gamma$-ray event. Each point represents the average charge density over a circular bin of width $5\,\rm m$ measured from the reconstructed shower core. The red line is an NKG fit with a fixed shower age of $s=1.6$. 
\emph{Right Panel:} Schematic of an air shower particle-front. The shower front is assumed to follow a hyperbolic shape to reconstruct the shower arrival direction ($\theta_{\rm R}, \phi_{\rm R}$). The slant height $S_\mathrm{max}$ of the shower maximum is determined assuming a spherical-shaped wavefront (see text for the details). Also shown is the shower plane which is perpendicular to the shower axis.
}}
\end{small}
\end{figure*}

\subsection{Reconstruction of the shower parameters}

The primary detectors (WCDs) record the complete footprint of the shower and are therefore used in the algorithms devoted to the reconstruction of the shower arrival direction and core position on the ground by an iterative strategy. 
The procedure which was implemented consists of a fitting loop where three functions are used iteratively to improve the reconstruction of the shower properties. 
We use the one-dimensional Nishimura-Kamata-Greisen (NKG) lateral charge distribution model \cite{Kamata,Greisen}, see figure~\ref{fig:wavefront} (left panel), to reconstruct the size ($N_{\rm pe}$) and the Moli\`ere radius ($r_M$) of the atmospheric showers while keeping the shower age parameter fixed at 1.6. Then we use the two-dimensional NKG model to reconstruct the shower core position ($X_{\rm R}, Y_{\rm R}$). Finally,
the hyperbolic shower front model \cite{LOFAR,LOPES}, see figure~\ref{fig:wavefront} (right panel), is used to reconstruct the arrival direction of the shower  ($\theta_{\rm R}, \phi_{\rm R}$), using the times and charges recorded in the WCDs. 
There are nine fitted parameters in total, combined in the different fits in the loop 
(see Table~\ref{table:reco}). 
Each WCD provides two data values (charge and time).  
So, a minimum number of six WCDs is required for the hyperbolic fit.  
But there is an interplay for the degrees of freedom between the fits in the iteration loop, 
such that we
 noted that after all cuts, there are no events with $N_{\rm WCD} \leq 8$, so we safely choose $N_{\rm WCD} \geq 8$ as a reasonable minimum.
In figure~\ref{fig:wavefront} (right panel), we also illustrate the definition of the \emph{shower plane} that we will use throughout the paper.

\begin{table}[t]
      \centering
      \small
      \begin{tabular}
      {lp{0.08\textwidth}p{0.45\textwidth}ccc}
        \hline
        \textbf{Variable} & \textbf{Used in} & \textbf{Definition} &\textbf{NKG} & \textbf{NKG} & \textbf{Hyp.}\\ 
         & \textbf{SEMLA} &  &\textbf{1D-fit} & \textbf{2D-fit} & \textbf{fit}\\ 
        \hline
        $s$ &  & Age of the shower & fixed &  fixed &  \\ 
        $r_{\rm M} \pm \Delta r_{\rm M} $ &  &  Moli\`ere radius & free & fixed &  \\ 
        $N_{\rm pe} \pm \Delta N_{\rm pe}$ & \checkmark & Size of the shower & free &  fixed &  \\ 
        \hline
        $X_{\rm R} \pm \Delta X_{\rm R}$ & \checkmark & Core position (X-coordinate) &  & free  & fixed \\ 
        $Y_{\rm R} \pm \Delta Y_{\rm R}$ & \checkmark & Core position (Y-coordinate) &  & free  & fixed \\ 
        \hline
        $l_{\rm R} \pm \Delta l_{\rm R}$ & \checkmark & Direction cosine with respect to X-axis &  &  & free  \\ 
        $m_{\rm R} \pm \Delta m_{\rm R}$ & \checkmark & Direction cosine with respect to Y-axis &  &  & free \\ 
        $d_{\rm 0}  \pm \Delta d_{\rm 0} $ &  & Perpendicular distance between the shower plane and the origin of the detector coordinates& &  &  free \\ 
        $\psi  \pm \Delta \psi$ &  & Cone angle of the asymptote with respect to the shower plane & &  &  free \\ 
        $a  \pm \Delta a$ &  & Perpendicular distance of the vertex of the hyperbola from the shower plane & &  & free \\ 
        
        \hline

    \end{tabular}
    
    \caption{
    \textbf{Variables from the shower reconstruction loop.} 
    The variables with the check marks in the second column are the ones that will be used in the SEMLA analysis.
    The prefix $\Delta$ represents the error associated with the fitted variable. }
       \label{table:reco}
\end{table}

\subsection{Rough evaluation of slant height of shower maximum}

In the development of air showers in the atmosphere, the depth at which the number of particles reaches its maximum is called the \emph{shower maximum}. 
The determination of that shower maximum is difficult in air-shower arrays.
For such detectors, no information is available on the shower development in the atmosphere, as only the footprint of particles reaching the array is recorded. But, a rough estimate of the slant height $S_{\rm max}$ of the shower maximum can be made assuming a spherical shape of the shower wavefront (see figure~\ref{fig:wavefront}, right panel). After the reconstruction of the shower core position and the arrival direction, $S_{\rm max}$ is determined by an independent fitting of a spherical shower wavefront to the recorded signals. 
Such an estimate assumes that all the air-shower particles are originated from a single point that is at a distance $S_{\rm max}$ on the shower axis.

\section{Basic principles of the development of SEMLA}
\label{sec:Analysis}

The SEMLA analysis strategy comes after the reconstruction of the shower properties and has the goal of optimising the detection of $\gamma$ rays. 
It comprises four stages, named A, B, C and D. 

In stage A, the reconstructed events with an extremely bad fit are cleaned using simple cuts. 
In stage B, poorly reconstructed events, i.e.\ events having large direction and core position reconstruction errors, are filtered out from good candidate events via a \emph{classification} procedure. 
In stage C, the $\gamma$-ray event candidates are discriminated from the proton-like events, again via \emph{classification}, and in stage D the primary energy of the final ``golden'' $\gamma$-ray events is reconstructed, this time via a \emph{regression} method. All the different stages in the SEMLA analysis are summarized in figure~\ref{fig_flow:ALTO_stages}.

In this section, we aim to introduce the basic principles and ideas behind the selection and the definition of the variables, and how we proceed in the machine learning techniques implemented in SEMLA, before going into the details of the analysis stages.

\begin{table}[t]
    \centering
    \small
    \begin{tabular}{p{0.25\textwidth}p{0.26\textwidth}p{0.26\textwidth}}
        \hline
        \textbf{Analysis stage} & \textbf{Signal} & \textbf{Background} \\ 
        \hline
        B (Filtering) & Well-reconstructed $\gamma$ rays & Poorly-reconstructed $\gamma$ rays  \\
        C (Proton suppression) & $\gamma$ rays  & Protons \\
        \hline
    \end{tabular}
    \caption{\textbf{Input sample definition for the SEMLA classification tasks.} Sample inputs for stages B and C of SEMLA.}
    \label{table:ChoiceOfMLPInput}
\end{table}

The classification and regression methods used in SEMLA take advantage of the feed-forward neural network  Multi-Layer Perceptron (MLP) class provided in the ROOT-TMVA package\footnote{ROOT version: 6.08.06 \& TMVA version: 4.2.0} \cite{ROOT, tmva}. The optimised training/testing options for the MLP procedures used in SEMLA are explained in section\ \ref{subsec:MLP_options}. 

The reconstructed shower properties listed in table~\ref{table:reco} and additional variables defined from the characteristics of the events are used as input variables for the training of the classification and regression methods. One of the input variables used in stage B classification is a Fisher discriminant, also implemented with the help of the ROOT-TMVA package.

\subsection{Datasets}

\label{sub_sec:data_work} 

Since the analysis involves machine learning techniques and definition of analysis cuts, the whole simulated data is equally split into two parts, hereby called the analysis implementation data set (\emph{ISet}) and the analysis performance data set (\emph{PSet}). 
The \emph{ISet} is used for the development of the analysis which is implemented with
the training and test procedures provided in ROOT-TMVA, while the \emph{PSet} is used to study the behaviour of the whole SEMLA analysis with an independent dataset in order to avoid that the evaluation of the analysis performance becomes too specialised on the \emph{ISet} which was used to define the cuts.

\subsection{Characteristics of the classification tasks}

\subsubsection{Input events}

In any binary classification method, the goal is to distinguish between 
a specific \emph{signal} and a certain number of cumulative sources of \emph{background}. 
In stages B and C of the SEMLA analysis, the definition of \emph{signal} and \emph{background} differs, as described below. In stage B, well-reconstructed $\gamma$-ray candidates are considered as \emph{signal} and poorly-reconstructed ones as \emph{background}. 
The $\gamma$-ray and proton events passing stage B are then used in stage C where the \emph{signal} is composed of $\gamma$ rays while the \emph{background} is composed of protons.
Table~\ref{table:ChoiceOfMLPInput} summarizes the input sample definitions for the classification stages B and C.

\subsubsection{Shower-size bins}
\label{sub_sec:binning}
The classification of events in $\gamma$-ray astronomy is highly dependent on the energy range under consideration.
The distributions of the discriminant variables implemented can vary substantially depending on the population of events considered.
Therefore, the classification performance is also highly dependent on the energy of the simulated $\gamma$-rays and protons. The \emph{background} rejection task can thus be facilitated by the definition of energy bins. The implemented reconstruction procedure does not provide the energy of the primary particle. Hence, in order to define the different energy ranges of the training/testing bins, we use the shower size $N_{\rm pe}$, see table~\ref{table:reco}, which is correlated with the true energy of the events, as shown in figure~\ref{fig:data_clean} (right panels). For this reason, in sections \ref{sub_sec:precut_performance} and \ref{sub_sec:stageC_performance}, a separate event classification will be carried out for each of three defined shower-size bins. 

The three bins are defined based on the value of $\log_{10}\left( N_{\rm pe}\right)$ such that we obtain almost equal number of events for training/testing procedure in all the bins and each bin has at least $2\times10^{4}$ events used for training process. 
In our train/test procedure, we select a set of input variables and the same set is used for the classification tasks in all three shower-size bins.

\subsubsection{Train/Test performance evaluation}
We study the performance of the train/test procedure by fixing the value of the \emph{signal} efficiency and evaluating the remaining amount of \emph{background} in the different bins. 
Our goal in SEMLA is to keep the \emph{background} efficiencies as low as possible, while requiring a constantly increasing \emph{signal} efficiency as a function of shower size or energy for the cut definition, consequently also ensuring a smooth increasing behaviour of the \emph{signal} effective area, as in \cite{new_analysis_strategy} section~$5.4$.

\subsection{Selection of relevant input variables}

\label{sub_sec:var_criteria} 

The selection and definition of variables for a multivariate analysis is a cumbersome and crucial process. In machine learning terminology, this process is known as \emph{feature engineering}.
Based on the physics of the atmospheric shower development and previous literature \cite{new_analysis_strategy, HAWC_NN, HAWC_Crab, hillas}, about 150 variables were initially considered for this work, but not all of them carry significant information. 

The possible variables are therefore screened based on the following criteria. We choose variables whose distribution shows some discrimination between \emph{signal} and \emph{background} events in every shower-size bin and that do not show a strong correlation or anti-correlation with others, so as to minimize the insertion of redundant information. 

The variables used in stage B and C are selected based on the above two conditions and scaled to the interval [-1,1] as recommended for MLP train/test procedures \cite{tmva, NN_book}.

For the analysis, we implement innovative variables based on the following: the Fisher discriminant, the predicted observables from the fit result, and the \emph{Hillas} parameters, see below.

The variables to be used in the analysis can be defined from the signals in the primary detectors (WCDs) and from the signals in the SD detectors.  
We summarize in table \ref{table:ChoiceOfDetectors} how we use the information acquired by the WCDs and SDs in the various steps of the analysis. 
WCDs provide significant information used consistently throughout all the analysis steps, while the SDs are used only in the trigger strategy, in the $\gamma$-proton separation procedure (stage C), and for the evaluation of the event energy (stage D).
The shower reconstruction is clearly based on the footprint seen by the primary detectors only, and the event cleaning (stage A) and filtering (stage B) procedures are conceived to select the best-quality subset of the reconstructed event sample through an essential use of the reconstruction errors.

\begin{table}[t]
    \centering
    \small
    \begin{tabular}{p{0.23\textwidth}p{0.06\textwidth}p{0.06\textwidth}}
        \hline
        \textbf{Analysis step} & \textbf{WCDs} & \textbf{SDs} \\ 
        \hline
        Trigger ``2/12'' & \checkmark  & \checkmark  \\
        Shower reconstruction  & \checkmark  &  \\
        SEMLA stage A    & \checkmark &  \\
        SEMLA stage B    & \checkmark  &  \\
        SEMLA stage C    & \checkmark  & \checkmark  \\
        SEMLA stage D    & \checkmark  & \checkmark  \\
        \hline
    \end{tabular}
    \caption{\textbf{Detectors used in the different analysis steps.} The information acquired by the WCDs is used at every analysis steps, while the information from the SDs is only used in the trigger phase and in the last stages of SEMLA.}
    \label{table:ChoiceOfDetectors}
\end{table}

\subsubsection{Defining linearly-combined input variables through the Fisher discriminant}
\label{sub_sec:Q_in}

The Fisher analysis is generally used for classification and regression problems with a given set of input variables, but in our analysis scheme, we use the method to linearly combine the information carried by two other observables into a new discriminant variable corresponding to the Fisher output. 
The Fisher algorithm looks for the hyperplane direction providing the best separation and smaller dispersion of the given event classes when projected upon. 
This is especially useful when two variables of different event classes exhibit a relation, where a significant portion of the background can be removed using a linear cut. 
We find that this is an easy and flexible method to define a new quantity varying smoothly over the two-dimensional phase-space of the initial two variables.
The Fisher output can then be used as one of the inputs for the MLP train/test procedures. 
An example of the application of this method is given by the definition of $Q_{\rm in}$ in section \ref{sub_sec:precut_variables}. 

\subsubsection{Predicted observables from the fit result}
\label{predictions}
The reconstruction models based on the fitted NKG and the hyperbolic functions, see section \ref{sec:simu_reco}, can be used to predict the signals in the array.
With the predicted signals, we can define predicted observables, in the same way we define the real observables, as described in \cite{new_analysis_strategy} section 4.3. 
When aiming for class separation, we find that the predicted observables carry additional discrimination power because they intrinsically contain the fit quality information.
Therefore, the predicted observables can be used alone in the train/test procedure or by comparing them to the real observables, as a significant difference between the predicted and the real observables is an indication that the function used in the fit is not satisfactorily representing a $\gamma$ ray. This allows to identify poorly-reconstructed $\gamma$-rays and protons.

Throughout the paper, we will refer to \emph{detected} signals and \emph{predicted} signals from the WCDs or SDs. 
A signal is \emph{detected} when in our simulations a WCD or a SD records an amplitude of more than $\SI{20}{\mV}$. A signal is instead \emph{predicted}, when, after shower reconstruction, we predict the integrated charge and time on a given WCD, according to the reconstructed shower properties. From the following section on, the variables obtained from the \emph{predicted} signals will be labelled with the subscript ``pred'' while those obtained from the \emph{detected} signals will have no label.

\subsubsection{Exploiting the Hillas parameters to define variables}

\begin{figure*}[t]
    \centering
    \includegraphics[width=0.32\textwidth, height = 0.32\textwidth] {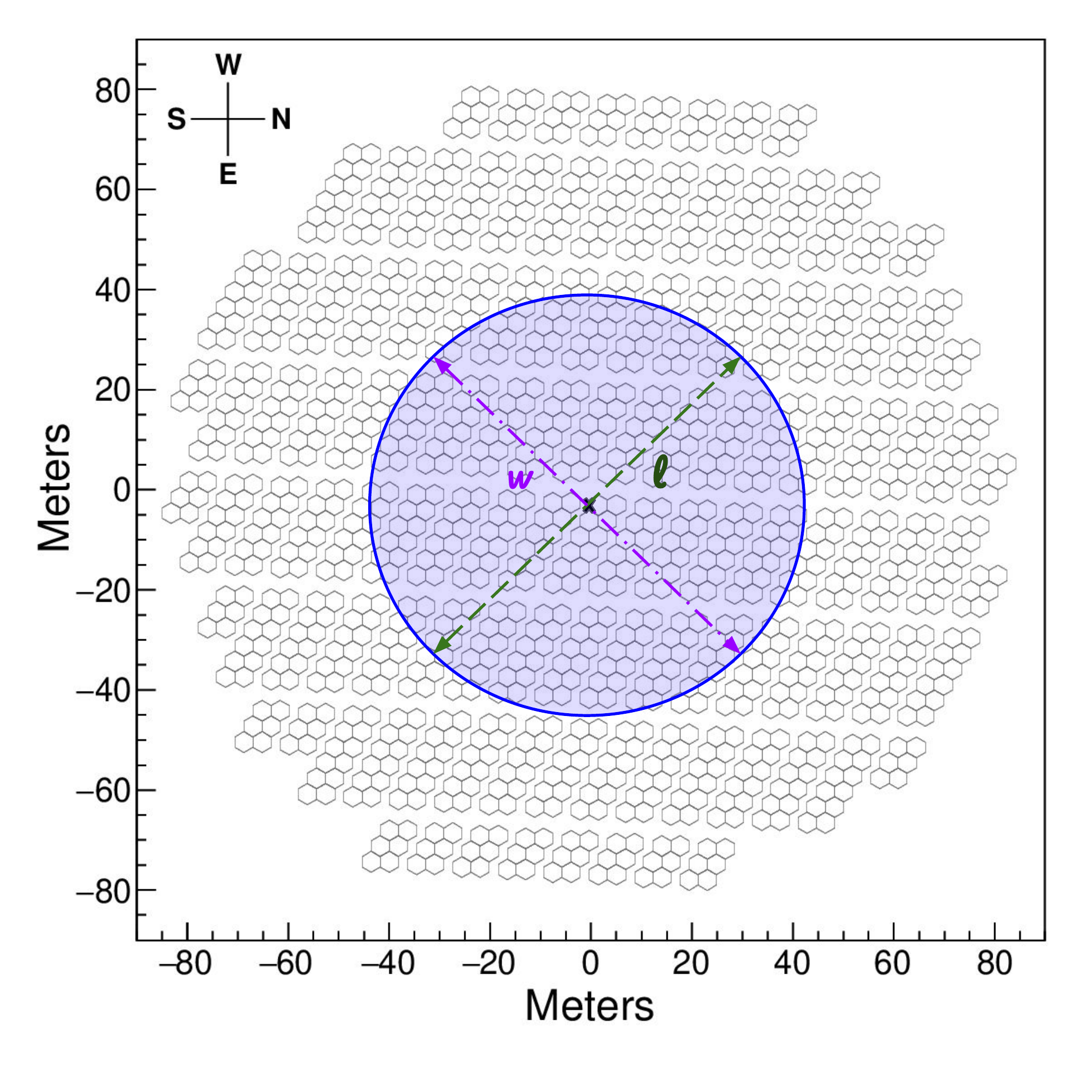}
    \hspace{0.1cm}
    \includegraphics[width=0.32\textwidth, height = 0.32\textwidth] {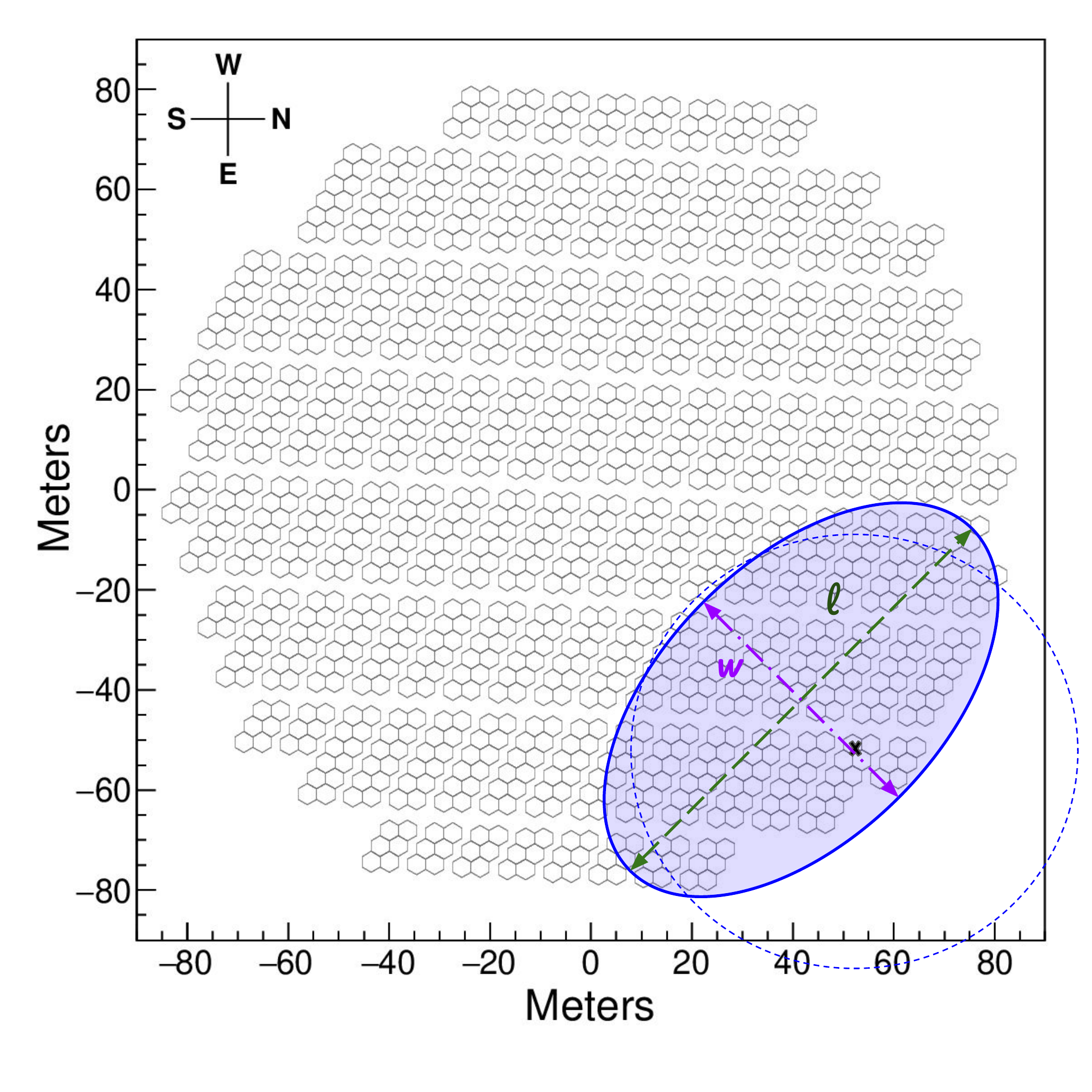}
    \hspace{0.1cm}
    \includegraphics[width=0.32\textwidth, height = 0.32\textwidth] {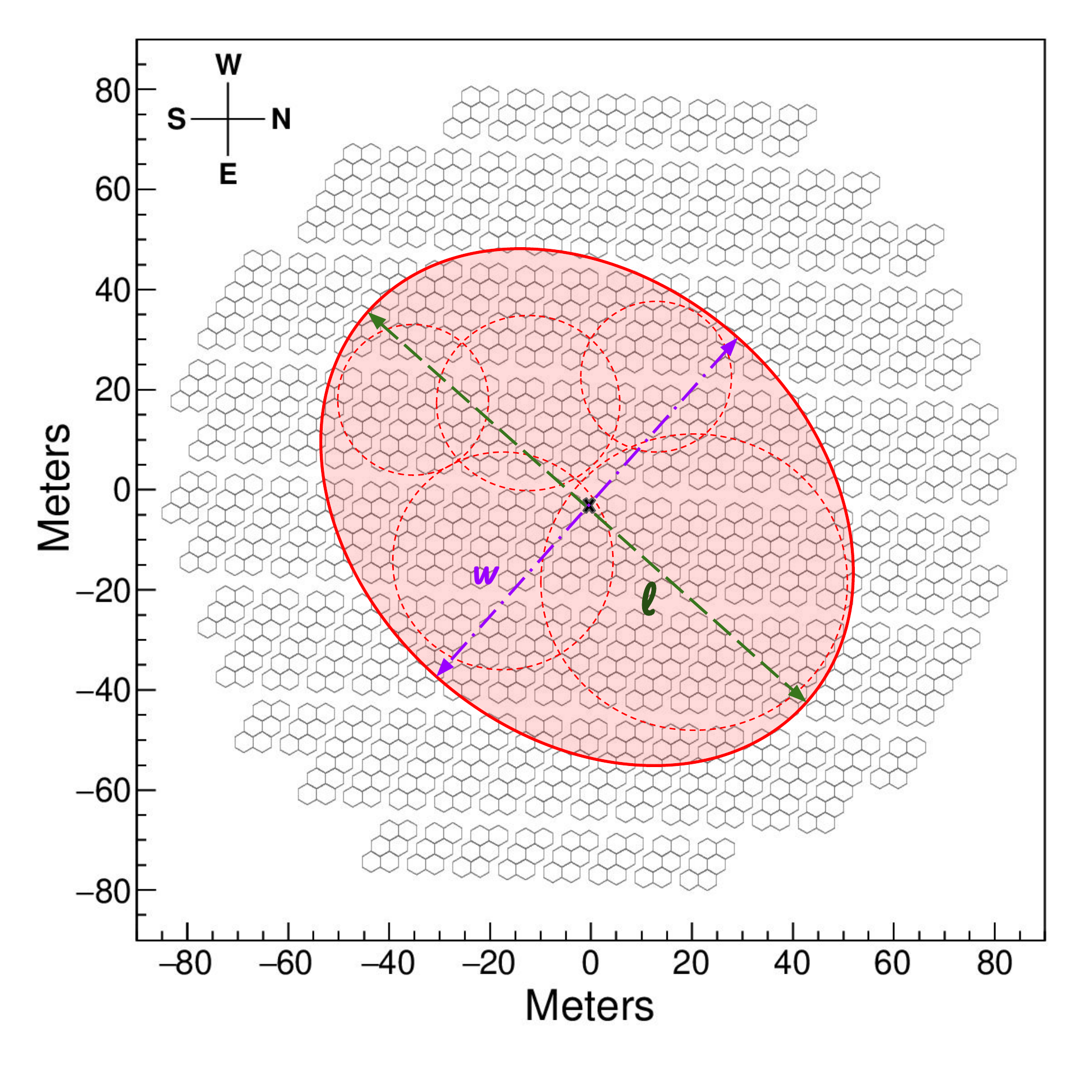}
    \caption{\small{\textbf{The Hillas-based variables in the framework of ALTO.} The variables $\ell,\,w$ represent the length and the width of the Hillas ellipse, respectively. 
    \emph{Left panel:} The boundary of a well reconstructed $\gamma$-ray shower footprint is shown in the ALTO shower plane. The X-mark indicates the true shower core, which is at the centre of the array. The blue circle is obtained from the Hillas parameters \cite{hillas} using triggered WCDs (not shown in the plots).
    \emph{Middle panel:} The case when the true core is at the edge of the array and therefore only the peripheral WCDs in the array are triggered. A large part of the event is lying outside the array, as shown by the dotted circle. This class of events is classified as poorly-reconstructed $\gamma$-rays and they have a larger value of $\ell$, the length of the major axis.
    \emph{Right panel:} The Hillas ellipse for protons (red ellipse) has a larger value of $\ell$, as it is elongated by WCDs triggered by muons and sub-showers (red dotted circles). For our classification tasks we found that the variable $\ell$ carries more information as compared to $w$ or the eccentricity, and hence it is used in the analysis.
    }}
    \label{fig:HillasDemo}
\end{figure*}

The Hillas parameters \cite{hillas}, based on the moments of the weights (pixel intensities) in the image plane, are commonly used in Imaging Atmospheric Cherenkov Telescopes for the characterization of the images generated by $\gamma$ rays. 
In our framework, we use them to characterise the footprint of the shower in the array, projected in the plane perpendicular to the reconstructed shower axis. 
Through the definition of Hillas-parameter-based variables, we aim to detect a difference in the footprints of an electromagnetic and hadronic shower on the detector array.
Indeed, for a well-reconstructed $\gamma$ ray, we would expect the footprint of the atmospheric shower to be very close to a circle. 
On the contrary, for a proton shower, the presence of muons and sub-showers can lead to significant deviations from a circular shape, and therefore we would expect the footprint to be more elongated, see figure~\ref{fig:HillasDemo}. We define a variable $\ell$ as the length of the Hillas ellipse obtained using the integrated charge of the signals detected in the WCDs, also see figure~\ref{fig:HillasDemo}. 

We also define $\ell_{\rm pred}$ with the NKG predictions calculated using the reconstructed shower, see section \ref{predictions}.
Indeed, when an event has its true core beyond or very close to the edge of the ALTO array, the shower footprint is seen only partially, leading to a more elongated image. 
By adding the information about the reconstructed core in the calculation of $\ell_{\rm pred}$, we enhance the differences between a well-reconstructed and a poorly-reconstructed event.

\subsection{Train/Test options for MLP procedure}
\label{subsec:MLP_options}
To perform the classification and regression tasks, we use the MLPs with the following configuration. The training/testing procedure is implemented using the \emph{ISet} data sample.

{\bf{Layers and neurons}}. The neural network used in this work consists of three layers. The first (input) and the second (hidden) layer contain an equal number of neurons, which is equivalent to the number of variables used in the training/testing procedure. The final layer consists of one neuron, which is the output of the neural network. The \texttt{tanh} function is used to activate the neurons. The number of hidden layers and the number of neurons in it were finalised after several tests to keep the same performance while reducing the computing time. 

{\bf{Algorithm}}. The training algorithm we use is the Broyden-Fletcher-Goldfarb-Shannon (BFGS) method with Bayesian extension \cite{tmva}.

{\bf{Statistics of train/test data in the \emph{ISet}}}. The bins in $\log_{10}\left(N_{\rm pe}\right)$ have their own number of signal and background.  
During the implementation of the analysis, we choose to keep a balance between the number of signal and background events. We aim not to specialize too much the performance of the analysis for a given type of $\gamma$-ray source, i.e. for a strong flaring $\gamma$-ray source as PKS~2155$-$304 \cite{PKS2155-304} versus for a faint ultra-soft-spectrum $\gamma$-ray source as KUV~00311$-$1938 \cite{KUV00311-1938}, for instance. With this in mind, we avoid having different statistics in the event populations that would need to be weighted according to the \emph{signal} searched for. 
Therefore, to ensure that the signal and background are given equal weight in terms of numbers of events in the training/testing process, we follow a method inspired by \cite{new_analysis_strategy} appendix A.5. The statistics of the events is therefore selected with the following simple procedure. 
First, we select the minimum between the number of the available signal $N_{\rm sig}$ and background $N_{\rm bkg}$ events in each bin
\begin{equation}
N_{\rm tot} = \min(N_{\rm sig},N_{\rm bkg}). 
\end{equation}
Then 60\% of $N_{\rm tot}$ is chosen randomly for the training process and the remaining 40\% are used in the testing process\footnote{We use the ROOT-TMVA \texttt{SplitMode=Random} procedure for this random choice of the events used for test or training within \emph{ISet}.} 
\begin{equation}
N_{\rm train} = 0.6 \cdot N_{\rm tot}, \;\;\;\;\; N_{\rm test}  = 0.4 \cdot N_{\rm tot}. 
\end{equation}

\section{Basic event cleaning in stage A}
\label{sec:data_clean}

The guiding idea of stage A is that we need to filter events -- based on observables -- for which the reconstruction has clearly failed to converge near the Monte Carlo true values, in order to facilitate the next steps of the analysis.
Removing outliers caused by errors in experimental equipment or failure of reconstruction procedures is a common step in machine learning \cite{ISL_book}. 
In addition, in a machine learning discrimination method, signal and background efficiencies need to be defined, resulting in only a partial rejection of the outlying events. For a complete rejection of outlying events in a specific region of phase space, simpler cuts are more efficient.

\begin{figure}[t]

\caption{\small \textbf{Stage A cleaning and stage B pre-cut successive effect}.  
The panels show the event distributions of two parameters with respect to the shower size $\log_{10}(N_{\rm pe})$.  \emph{Left panels:} The distributions for the experimentally-observable logarithmic fractional error in the reconstructed shower $N_{\rm pe}$ over the fitted value of ${N_{\rm pe}}$, $\sfrac{\Delta N_{\rm pe}}{N_{\rm pe}}$.  \emph{Right panels:} The distributions for the true energy, $\log_{10}(E_{\rm T})$.
The upper three rows of panels show the successive effect  -- for the simulated $\gamma$ rays -- starting from the reconstructed events for which $N_{\rm WCD} \geq8$ \emph{(first row)}, then after stage A (event cleaning, section\ \ref{sec:data_clean}) \emph{(second row)}, and lastly after the stage B pre-cuts \emph{(third row)} (section\ \ref{sec:select_best}).  The lowest panels show these distributions 
for the simulated protons after stage B, for comparison.
The purple dot-dashed line on the upper-left panel shows the ``Blob-cut'' as defined in section\ \ref{sec:data_clean}, which removes both the ``blob'' events and those with excessive fractional error in the $N_{\rm pe}$ fit. 
The brown dot-dashed lines in the right panels indicate the main and secondary branch mentioned in the text. The plots are obtained with the \emph{PSet} data sample.
}

{\includegraphics[width=1.05\textwidth,trim={30pt 0 0pt 30pt},clip]{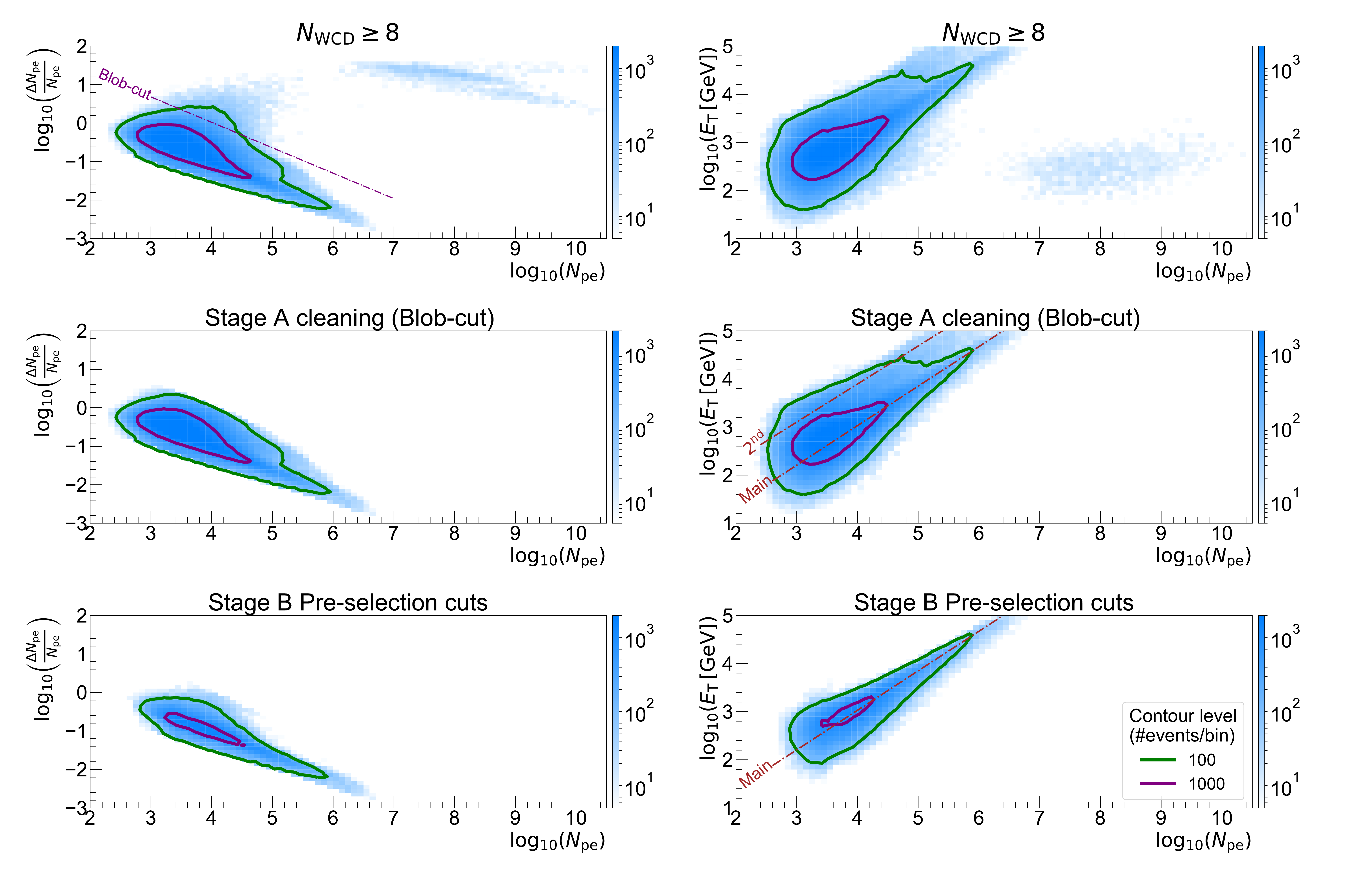}}
\\
\hfill\includegraphics[width=1.05\textwidth,trim={30pt 0 0pt 630pt}]{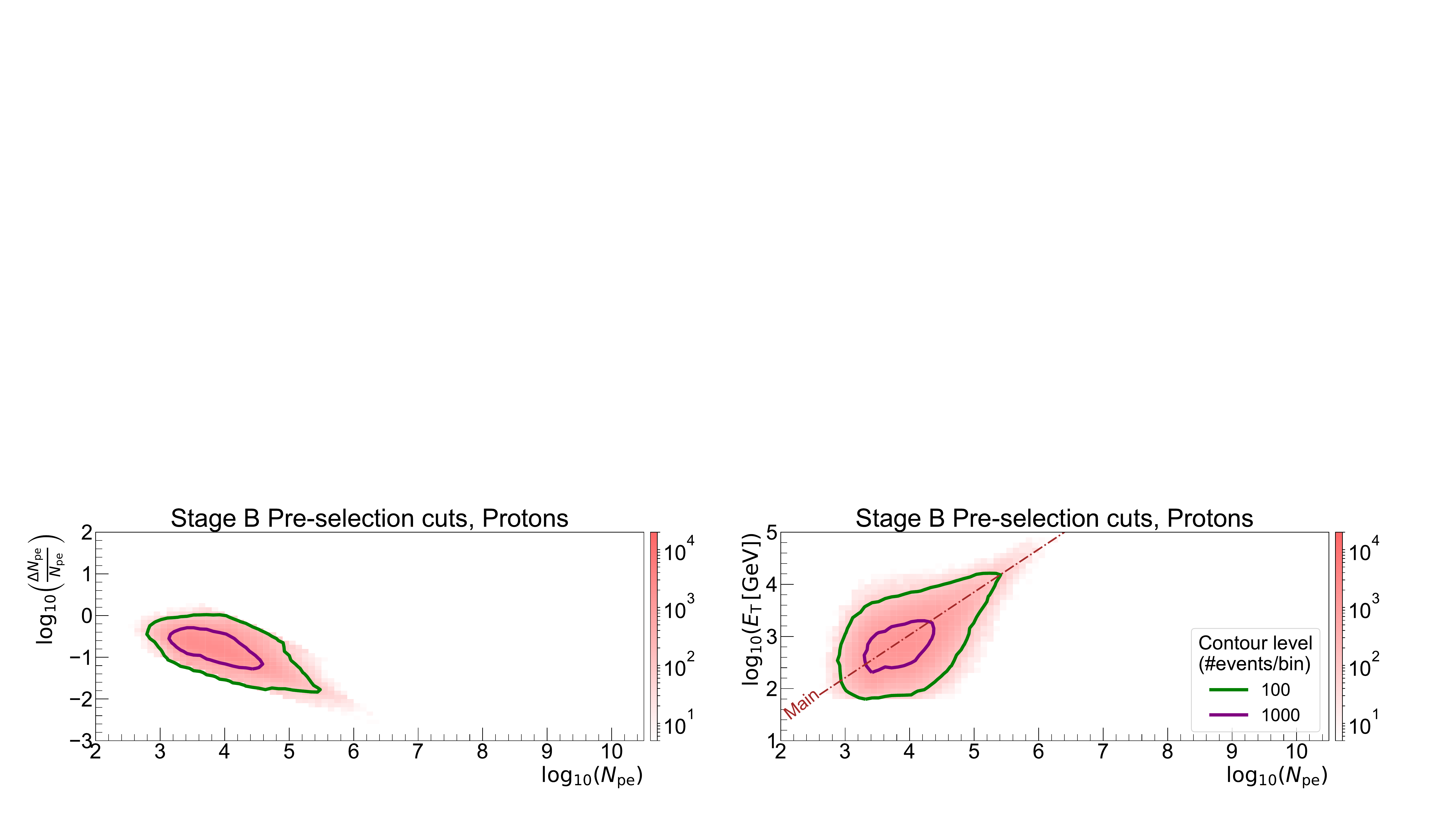}
\label{fig:data_clean}
\end{figure}

We study the relation between the true energy $E_{\rm T}$ of the events and total number of detected photo-electrons $N_{\rm pe}$ in the ALTO WCDs.
Figure~\ref{fig:data_clean} (upper-right panel) shows this relation.
For most of the events,  $E_{\rm T}$ is correlated with $N_{\rm pe}$, but there is a distinct \emph{blob} of low-energy events with very high values of reconstructed $\log_{10}(N_{\rm pe}) \sim$ 7--9. 
It is evident that, except for the events showing a strong linear correlation between $\log_{10} (E_{\rm T})$ and $\log_{10} (N_{\rm pe})$, the rest are events for which the reconstruction procedure has failed. 

The events in the blob --- with their high value of $N_{\rm pe}$ --- likely result from  a very flat fit of the lateral distribution of the NKG function, which can occur, for instance, when a small number of detectors are triggered. The NKG flattening effect is also reflected through the error on $N_{\rm pe}$ (see  figure~\ref{fig:data_clean}, upper-left panel), for the same events with high $\log_{10}(N_{\rm pe})$ values.

The preliminary step to remove these wrongly reconstructed events is to cut based on the fitted variable errors. After requiring that the fit has properly converged, we apply the following \emph{Blob-cut} to all events.
    
{\textbf{Blob-cut}}: By studying the relation between the fractional error in $N_{\rm pe}$ and $N_{\rm pe}$ (figure~\ref{fig:data_clean}, upper-left panel) we define a cut to remove both the blob events and those with an excessive fractional error in the fit of $N_{\rm pe}$, as follows:
\begin{equation}
            \centering
                \log_{10}\left( \frac{\Delta N_{\rm pe}}{N_{\rm pe}}\right)  < [-0.66\cdot \log_{10}(N_{\rm pe})] + 2.66
\end{equation}

The overall effect of the cuts applied in stage A is shown in figure~\ref{fig:data_clean} (second row). A clear linear correlation between $\log_{10} (N_{\rm pe})$ and $\log_{10} (E_{\rm T})$ is visible in the cleaned events. This correlation is indicated as ``Main'' in figure~\ref{fig:data_clean} (second row, right panel). However, an additional parallel secondary branch of events at higher energy is also noticeable, indicated as ``$\rm 2^{nd}$''.
We have determined that the events in the secondary branch have a true impact parameter outside the ALTO array and are therefore triggering only some of the edge detectors. 
Since in this case the shower footprint is only partly observed, this leads to an underestimate of $N_{\rm pe}$ and the creation of the secondary branch. 
These events will be removed by stage B of the analysis, as can be seen by the solid contours in figure~\ref{fig:data_clean} third row, right panel.

\section{Filtering of poorly reconstructed events in stage B}
\label{sec:select_best}

As mentioned in section \ref{sec:simu_reco}, point-like $\gamma$ rays and diffuse protons are simulated with a maximum true impact parameter of $130\,\rm{m}$ and $184\,\rm{m}$ respectively, therefore in an area larger than the detector array.
Depending on the energy of the primary, the events with a true impact parameter at the edge or beyond the array ($b_{\rm T}\geq\rm80\,\rm{m}$), can still eventually trigger 8 or more WCDs ($N_{\rm WCD} \geq8$).
These \emph{peripheral} events are always reconstructed with an impact parameter ($b_{\rm R}$) within the array, therefore the evaluation of the impact parameter by the reconstruction procedure is wrong. 
The secondary branch in the right panels of figure~\ref{fig:data_clean} mainly comprises this category of events.

The focus of the stage B in SEMLA is to remove poorly-reconstructed events from the whole population, which includes also the events in the secondary branch, through an MLP classification procedure. In this step we only use $\gamma$-ray simulations as input for the training, because the cut must be optimized for the detection of well-reconstructed $\gamma$ rays.

\subsection {Input sample definition}
\label{sub_sec:precut_input_def}

For the definition of the input samples we take advantage of the Monte Carlo information and the reconstruction quality of the events. 
A $\gamma$-ray event is said to belong to the \emph{signal} of well-reconstructed events if:
\begin{enumerate}[(i)]
    \item the true impact parameter is within the array ($b_{\rm T} \leq $ 80 m); 
    \item the angular distance between the true $(\theta_{\rm T},\phi_{\rm T})$ and the reconstructed $(\theta_{\rm R},\phi_{\rm R})$ direction is less than 1.5$^\circ$;
    \item the true and reconstructed shower cores are a maximum of $20\,\rm m$ apart: 
    \begin{equation}
        \sqrt{(X_{\rm T}-X_{\rm R})^2+(Y_{\rm T}-Y_{\rm R})^2}\leq 20 \,\rm m;
    \end{equation}
\end{enumerate}
$\gamma$-ray events not satisfying these conditions are considered as \emph{background} for this stage. 

\subsection{Input variables}
\label{sub_sec:precut_variables}

\begin{figure*} [t]
    \centering
    
    \includegraphics[width=\textwidth]{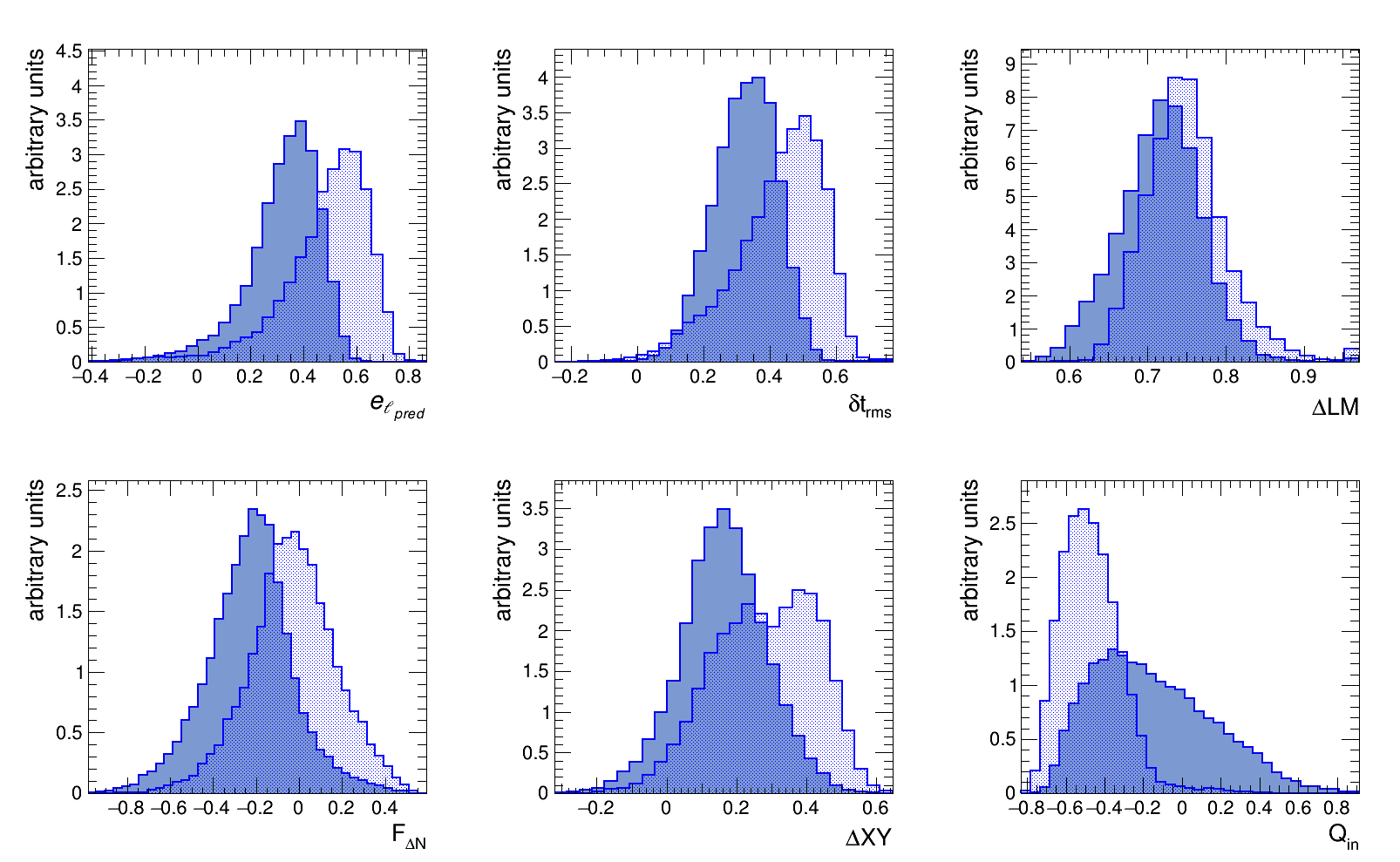}
    \caption{\small{\textbf{Input variables for stage B in bin B3.} Distributions of discriminant variables used in the MLP pre-selection training to separate poorly-reconstructed events (blue, dotted shading) from well-reconstructed events (blue, solid shading). The definitions of the selected variables can be found in section \ref{sub_sec:precut_variables}. The plots are obtained with the \emph{ISet} data sample.}}
    
    \label{fig:precut-variables}
\end{figure*}

The six variables chosen for this MLP selection procedure are selected based on the criteria explained in section \ref{sub_sec:var_criteria}, and their distributions for well-reconstructed and poorly-reconstructed $\gamma$-ray events can be seen in figure~\ref{fig:precut-variables}. 
These are:  $e_{\ell_{\rm pred}}$, $\delta t_{\rm rms}$, $\Delta LM$, $F_{\Delta N}$, $\Delta XY$ and $Q_{\rm in}$.

    \bm{$e_{\ell_{\rm pred}}$}: The predicted integrated charge in the triggered water tanks can be obtained using the NKG function and the the reconstructed shower information $N_{\rm pe}$, core position $(X_{\rm R},Y_{\rm R})$ and $r_{\rm M}$. 
    With the predicted integrated charges on each WCD, one can calculate the Hillas ellipse obtaining the length of the ellipse $\ell_{\rm pred}$, and then its logarithm
    \begin{equation}
        e_{\ell_{\rm pred}}=\log_{10}(\ell_{\rm pred})
    \end{equation}
    Though both $\ell$ (defined with the detected charges) and $\ell_{\rm pred}$ can characterize the shower footprint and event containment in the array, $\ell_{\rm pred}$ additionally contains the information about the fit quality (through the predicted charges) and is therefore more discriminant in the task of selecting well reconstructed events.
    The distribution of $e_{\ell_{\rm pred}}$ shows that for a well-reconstructed event the value of $e_{\ell_{\rm pred}}$ is smaller than for a poorly reconstructed event, as shown in figure~\ref{fig:precut-variables} (upper left panel).
    
    \bm{$\delta t_{\rm rms}$}: Similar to the first variable, we calculate the predicted times in the triggered water tanks by using the hyperbolic shower-front model. We define $\delta t_{\rm rms}$ as
    \begin{equation}
            \delta t_{\rm rms} = \log_{10} \left (\sqrt{\frac{1}{N_{\rm WCD}} \sum_{i=1}^{N_{\rm WCD}}({t^{\rm det}_{i}}-{t^{\rm pred}_{i}})^{2}} \right )
        \end{equation}
    where $t^{\rm det}_{i}$ is the time obtained in a WCD, and $t^{\rm pred}_{i}$ is the time predicted in a WCD from the hyperbolic shower front fit. For a well-reconstructed event the distribution of the time residuals $({t^{\rm det}_{i}}-{t^{\rm pred}_{i}})$ has a smaller root-mean-square (RMS) value than a poorly-reconstructed event.
    
    \bm{$\Delta LM$}: Based on the straightforward reasoning that well-reconstructed events have a smaller error in the reconstructed directional cosines with respect to poorly-reconstructed ones, the variable is defined as
        \begin{equation}
            \Delta LM = \log_{10} [\left (\Delta l_{\rm R}\right) ^2 + \left( \Delta m_{\rm R} \right )^2 ] 
        \end{equation}
    where $\Delta l_{\rm R}$ and $\Delta m_{\rm R}$ are the fit errors in the reconstructed directional cosines. 
    
    \bm{$F_{\Delta N}$}: The logarithmic value of the fractional error in the reconstructed size of the shower, $F_{\Delta N}$, is defined as
        \begin{equation}
            F_{\Delta N} = \log_{10}\left( \frac{\Delta N_{\rm pe}}{N_{\rm pe}}\right) 
        \end{equation}
    and helps removing poorly-reconstructed events, as for these the error in $N_{\rm pe}$ is larger than $N_{\rm pe}$.

    \bm{$\Delta XY$}: Given the fit errors in the reconstructed core position $\Delta X_{\rm R}$ and $\Delta Y_{\rm R}$, as provided by the NKG function, we can define
        \begin{equation}
                \Delta XY = \log_{10} [ \left (\Delta X_{\rm R}\right) ^2 + \left( \Delta Y_{\rm R} \right )^2 ]. 
        \end{equation}
    Similarly to $\Delta LM$, poorly-reconstructed events have a larger $\Delta XY$.
   
    \bm{$Q_{\rm in}$}: A close examination of the relation between the variable $b_{\rm R}$ and $\log_{10}(N_{\rm pe})$ for well- and poorly-reconstructed events shows that low-energy events with smaller impact parameter are well-reconstructed, while those with a high impact parameter are poorly-reconstructed. For example, a $\gamma$-ray event with $E_{\rm T} = 500\,\rm GeV$ and an impact parameter of $90\,\rm m$ triggers 11 tanks ($N_{\rm pe} = 300$) at the edge of the array. 
    These events are poorly fitted by the NKG function and the reconstructed impact parameter is found to be $\rm 75\,m$, the underlying reason being that only the charges in the tail of the distribution are observed. 
    For the same energy but with the true impact parameter of $\rm 48\,m$, 13 tanks ($N_{\rm pe} = 400$) are triggered close to the centre of the array. This event is well reconstructed with an impact parameter of $\rm 46\,m$. 
    This shows that with similar energy and similar number of WCDs triggered, the quality of reconstruction highly depends on the impact parameter. 

    $Q_{\rm in}$ is developed with the purpose of detecting poorly reconstructed core positions as a function of the shower size, and we construct the variable with the help of the Fisher discriminant. We define $Q_{\rm in}$ as 
    \begin{equation}
        Q_{\rm in} = F_0 +  F_1 \log_{10}(N_{\rm pe}) + F_2 b_{\rm R}
    \end{equation}
    where the coefficients $F_0$, $F_1$ and $F_2$ obtained from the Fisher algorithm are 0.46, 0.74 and $-$0.65 respectively. 
    The well-reconstructed $\gamma$-ray \emph{signal} and poorly-reconstructed \emph{background} samples that we use in the Fisher procedure are the same as those defined in section \ref{sub_sec:precut_input_def}. 
    The Fisher train/test options that we use for this task are the default ones provided in \cite{tmva}.
    Summarizing, this variable helps in removing many poorly-reconstructed low-energy events close to the edge of the array.
    
\subsection{Performance of stage B} 
\label{sub_sec:precut_performance}

 \begin{table} [t]
    \centering
    \small
\begin{center}

        \begin{tabular}{cccccccc}
        \hline 
        
        \textbf{Bin}  & \bm{${\rm log}_{10}\left(N_{\rm pe}\right)$}  & \textbf{Required} &  \bm{$N_{\rm train}$} & \bm{$N_{\rm test}$} & \multicolumn{1}{c}{\textbf{Obtained}} & \multicolumn{1}{c}{\textbf{Cut}}& \multicolumn{1}{c}{\textbf{Selected events'}} \\ 
        
        \textbf{name} & \textbf{range} &  \textbf{signal} &    &  & \multicolumn{1}{c}{\textbf{\bm{$\varepsilon_{\rm BG}$} (\%) on}} & \multicolumn{1}{c}{\textbf{value on}}&  \multicolumn{1}{c}{\bm{${\rm log}_{10}\left(E_{\rm T}\right)$}}\\
        
        &  & {\textbf{\bm{$\varepsilon_{\rm Sig}$} (\%)}} & & & \textbf{test data} & \textbf{test data}&  \multicolumn{1}{c}{\textbf{mean \bm{$\pm$} sigma}}\\
        
        \hline
        
        B1 & 1.00 -- 3.54 & 50 & 22840  & 15228 & $9.83\pm0.25$ & 0.71 & $2.61\pm0.28$\\
        B2 & 3.54 -- 3.94 & 70 & 22633  & 15089 & $7.36\pm0.22$ & 0.71 &  $2.80\pm0.27$\\
        B3 & 3.94 -- 7.00 & 90 & 21673 & 14450 & $4.54\pm0.18$ & 0.63 &  $3.37\pm0.46$\\
        \hline

        \end{tabular}
    \end{center}

    \caption{\small{\textbf{Stage B performance.} MLP bins showing the efficiency of well-reconstructed $\gamma$-ray \emph{signal} and poorly-reconstructed $\gamma$-ray \emph{background}, the number of training and test events, the cut values on test data and the mean and sigma of $\log_{10}(E_{\rm T})$. Note that the cut value varies for different shower-size bins. An equal number of events is used for \emph{signal} and \emph{background} training, given by $N_{\rm train}$, and similarly for $N_{\rm test}$. 
    We chose the values of the \emph{signal} efficiency ($\varepsilon_{\rm Sig}$) considering that the effective area increases as a function of the energy and that we wanted to keep the \emph{background} efficiency ($\varepsilon_{\rm BG}$) $\leq10\%$. The error in the background efficiency is the statistical one. The statistics are obtained with the \emph{ISet} data sample.}}

    \label{table:precut_bins}
\end{table}

\begin{figure*}[t]
{\includegraphics[width=0.48\textwidth]{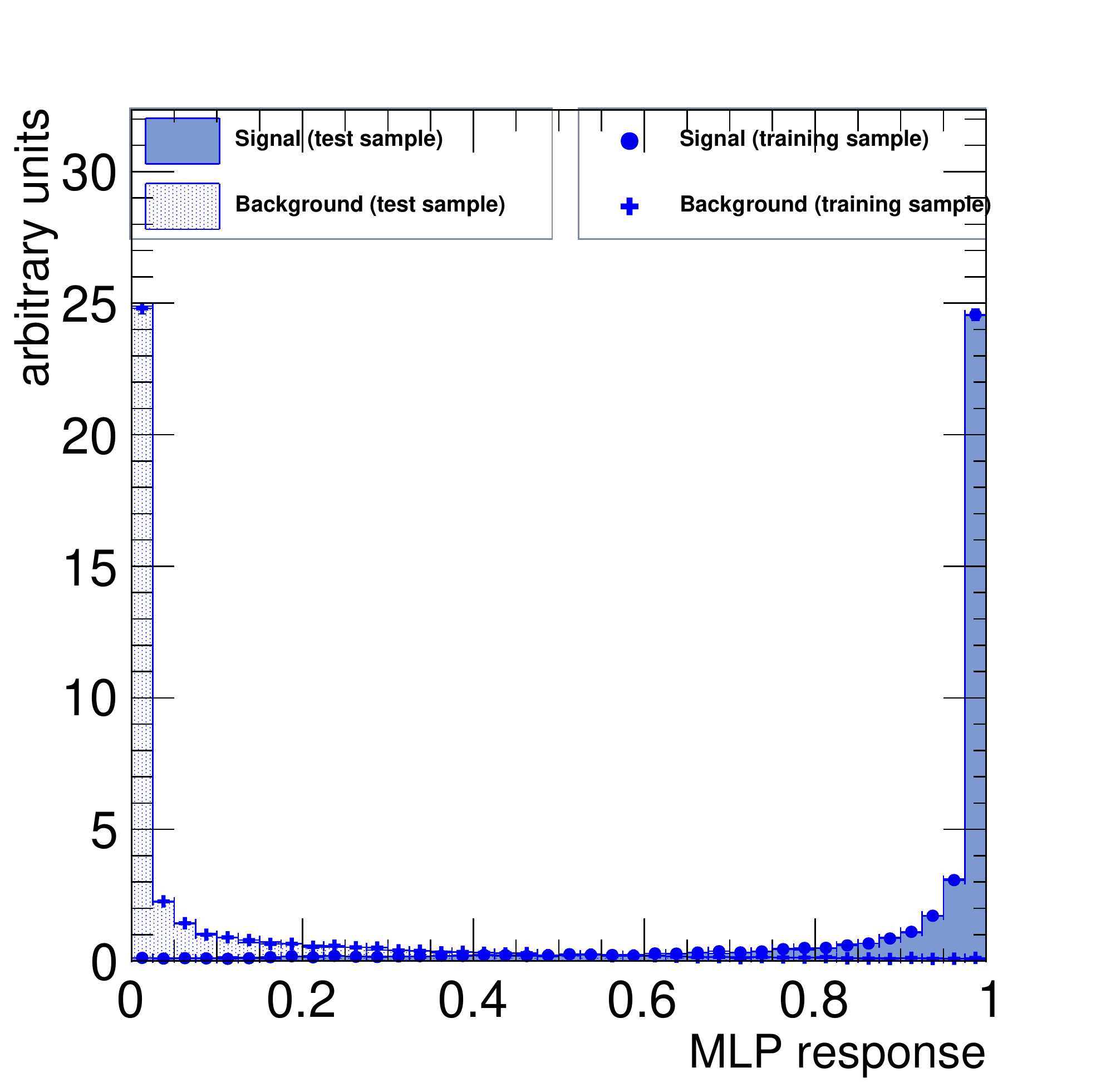}}
{
    \caption{
        \small{
            \textbf{Stage B MLP response.} The MLP classifier response for the stage B pre-selection for a signal of well-reconstructed $\gamma$ rays versus the background of poorly-reconstructed ones, for the highest shower-size bin, B3.  The plot is obtained with the \emph{ISet} data sample.}
            }
    \label{fig:StageB_MLP_response}
}

\end{figure*}

As mentioned above, the test/train procedure of stage B is performed in shower-size bins.
For stage B, the \emph{signal} is composed of the  well-reconstructed events, as defined in section~\ref{sub_sec:precut_input_def}, versus the \emph{background} of the remaining poorly-reconstructed events. The MLP train/test response for the bin B3, with highest $N_{\rm pe}$, is shown in figure~\ref{fig:StageB_MLP_response}. We define the cut based on the MLP response variable, which ranges from 0 to 1 according to the probability of an event being well-reconstructed, so as to obtain a pre-defined \emph{signal} efficiency.    

Table~\ref{table:precut_bins} shows the definition of the training bins, the number of events used in training/testing procedure, the required well-reconstructed $\gamma$-ray \emph{signal} efficiencies, the corresponding \emph{background} efficiency of poorly-reconstructed events for test data,  and the cut values obtained. 
Given our \emph{signal} efficiencies and our train/test performance, the fraction of poorly-reconstructed $\gamma$-ray events decreases with increasing energy. 
The training/testing procedure is applied independently in the three shower-size bins, resulting in three different cut values. The three cuts obtained in stage B, see table~\ref{table:precut_bins}, are applied to both the simulated $\gamma$-rays and protons.

\section{Proton suppression in stage C}
\label{sec:gamma-hadron}
Following the suppression of poorly-reconstructed events, we are left with the crucial part of the analysis, which is $\gamma$/proton separation. 
In this stage, we add the information acquired through the SDs, designed for muon tagging.
This analysis step also employs an MLP classification training/testing method with different input variables
and definition of \emph{signal} and \emph{background} with respect to the one used
in stage B.
After applying the stage B cuts, all remaining $\gamma$-ray events are taken as the \emph{signal} for stage C, and the protons remaining are taken as the \emph{background}. 
For this stage, fifteen variables are chosen for training/testing process. We reuse the five variables defined for stage B, namely $e_{\ell_{\rm pred}}$, $\delta t_{\rm rms}$, $\Delta LM$, $F_{\Delta N}$,  $\Delta XY$, as well as the reconstructed $b_{\rm R}$, $N_{\rm pe}$ and $S_{\rm max}$.
In addition, we use the length of the major axis of the Hillas ellipse defined from the detected charge, $e_{\ell}$ defined as
    \begin{equation}
        e_{\ell}=\log_{10}\left (\ell \right)
    \end{equation}
 and two basic variables: the total number of triggered WCDs, $N_{\rm WCD}$, and the total charge detected in an event, $Q_{\rm WCD}$. 

For this stage we increase the number of variables by exploiting further the WCD information ($\delta q_{\rm mean}$, $\delta q_{\rm rms}$) and by using the SD information for the first time in the analysis ($Q_{\rm SD}$ and $\log_{10}(N_{\rm SD}+10)$), see below. 

\subsection{Additional WCD-based variables}

Starting from the reconstructed shower, for each event, we define the charge residuals as $({q^{\rm det}_{i}}-{q^{\rm pred}_{i}})$, where $q^{\rm det}_{i}$ is the charge detected in a WCD, and $q^{\rm pred}_{i}$ is the charge predicted in a WCD from the NKG fit.  
From the charge-residuals distribution we calculate the mean value $\delta q_{\rm mean}$ and the root-mean-square $\delta q_{\rm rms}$, since we expect that both distributions should show a significant difference between $\gamma$ rays and protons. 

\begin{figure*}[t]
    \centering
    
    \begin{subfigure}{\textwidth}
        \centering
        \includegraphics[width=\textwidth]{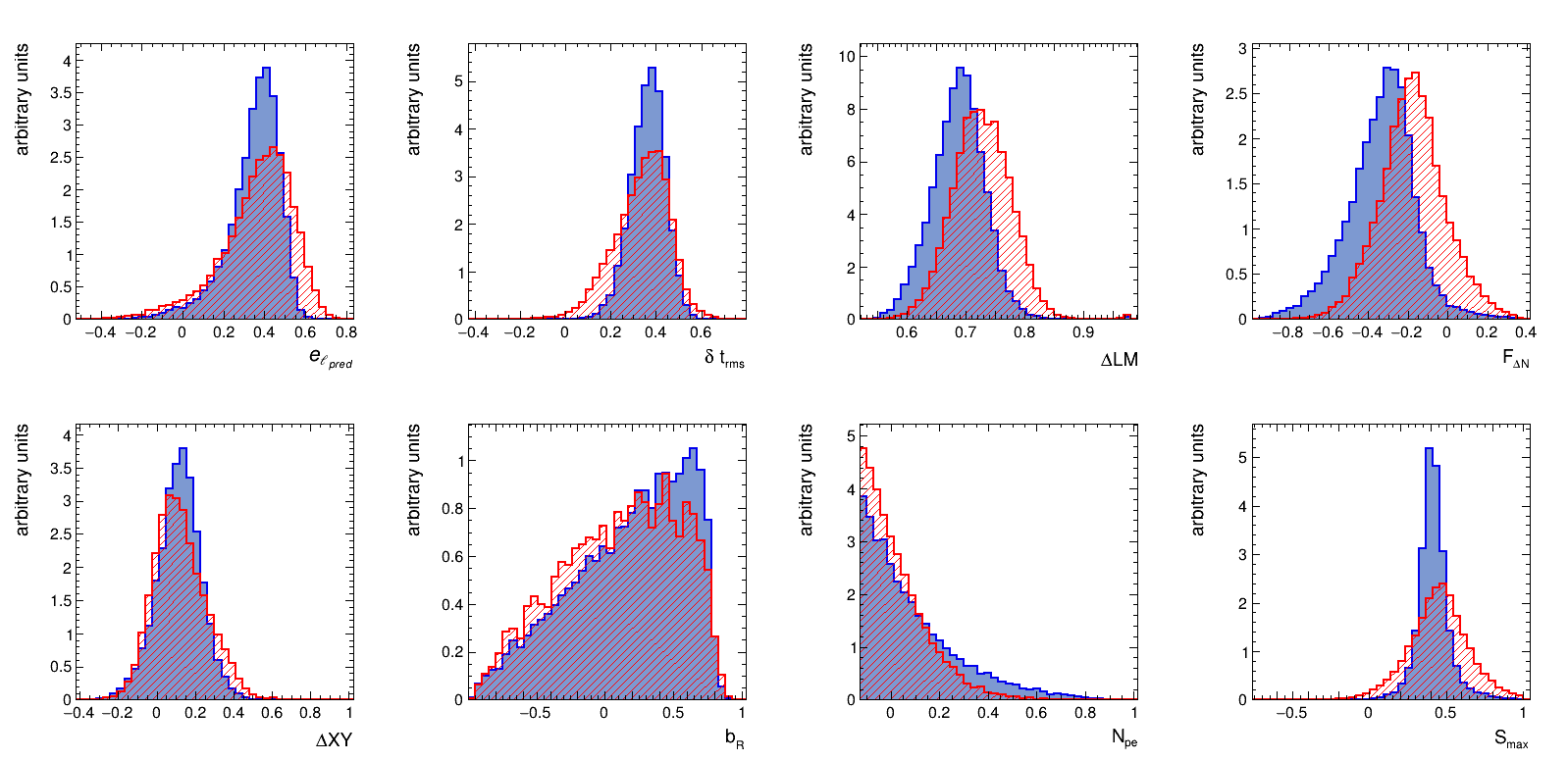}
    \end{subfigure}
    
    \begin{subfigure}{\textwidth}
        \centering
        \includegraphics[width=\textwidth]{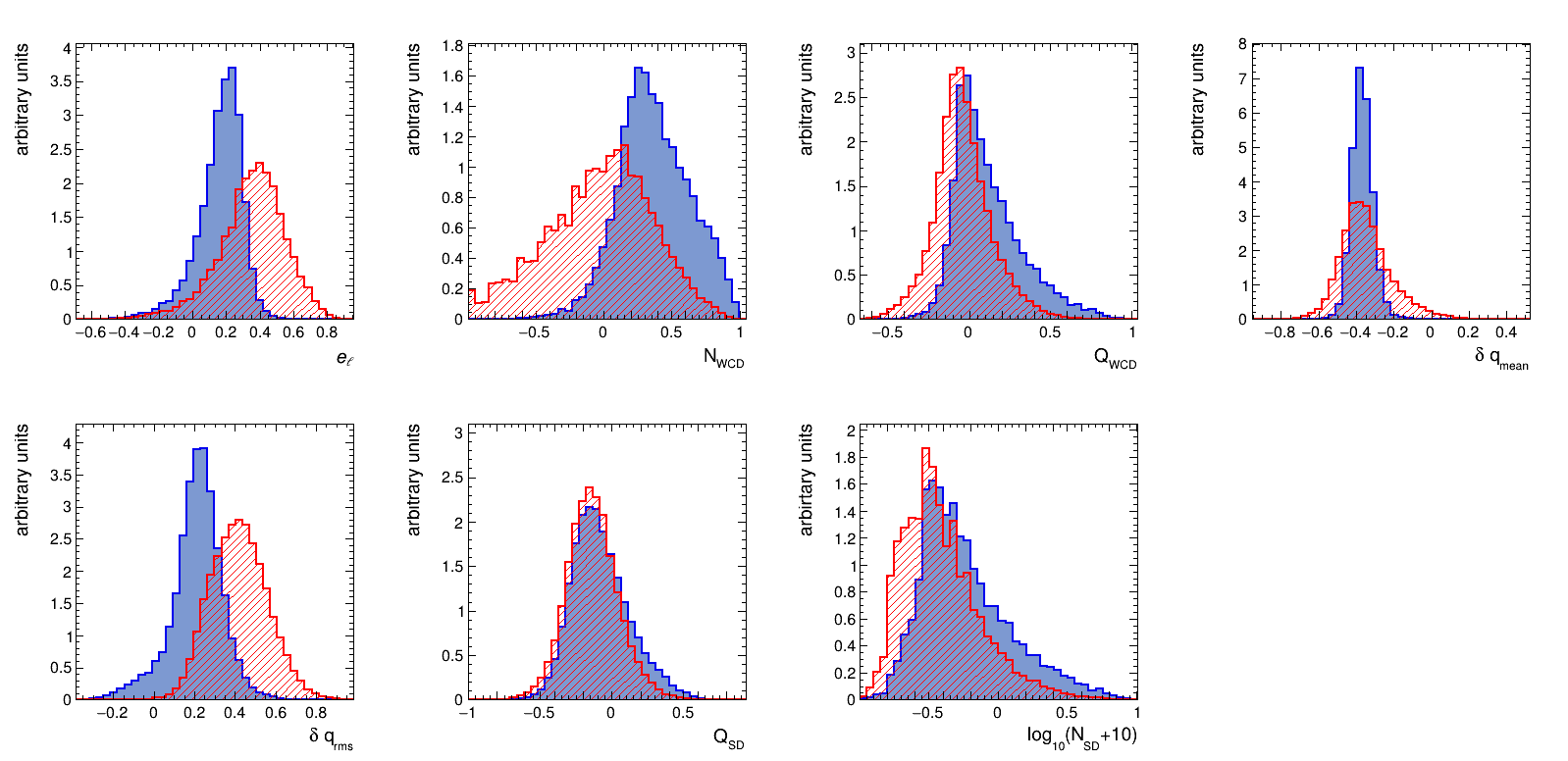}
    \end{subfigure}
    
    \caption{\small{\textbf{Input variables for stage C in bin number C3.} Input variables used for the $\gamma$/proton separation MLP to extract the $\gamma$ ray \emph{signal} (blue,  shaded) from the proton \emph{background} (red, dashed). The variables are defined in section\ \ref{sec:gamma-hadron}. The plots are obtained with the \emph{ISet} data sample.}}
    \label{fig:StageC_variables}
\end{figure*}

\begin{figure*}[t]
{\includegraphics[width=0.45\textwidth]{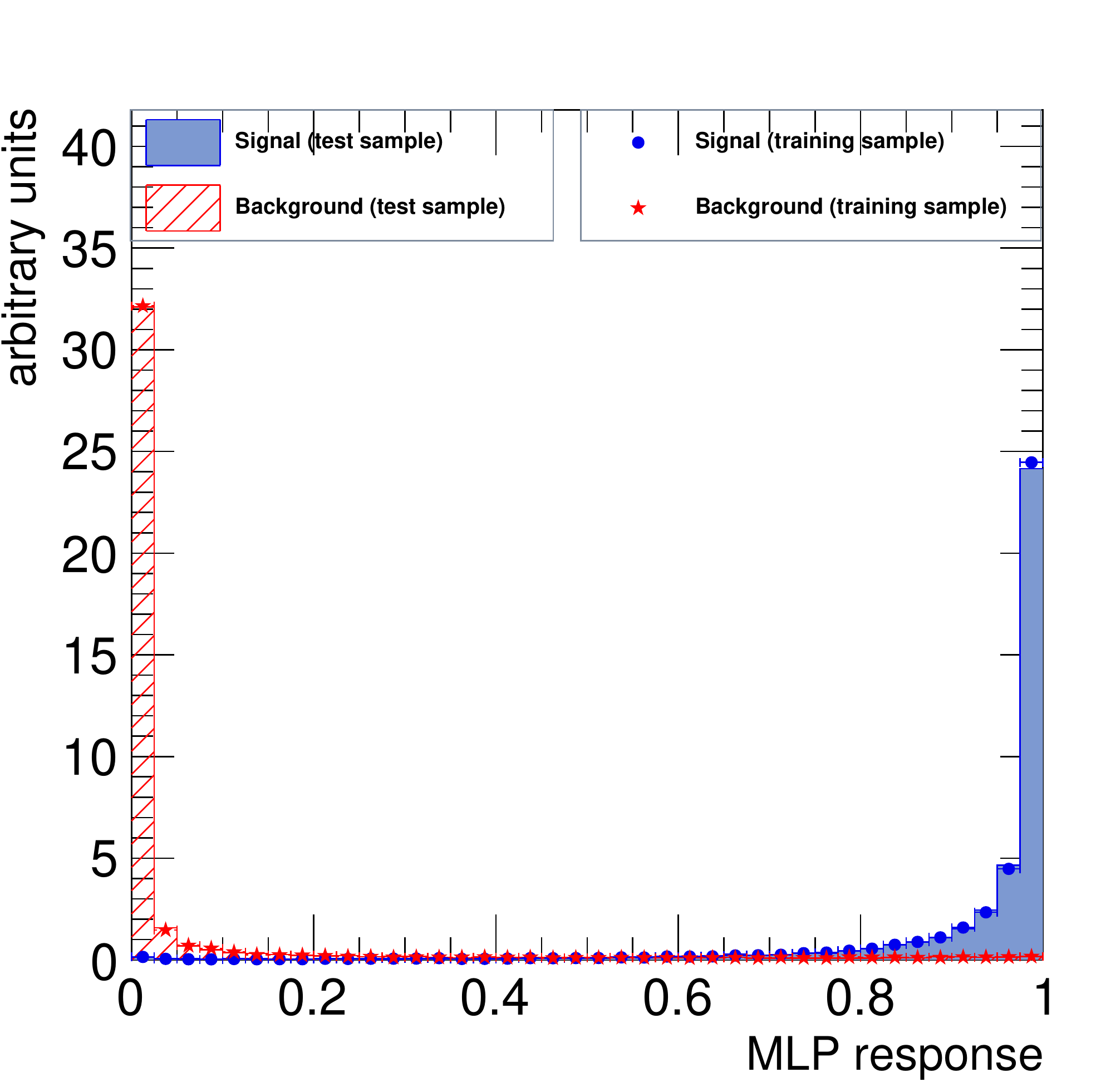}}
{\caption{\small{\textbf{Stage C MLP response.} MLP classifier response for the stage C selection of the $\gamma$-ray signal versus the proton background for the highest shower-size bin, C3.  The plot is obtained with the \emph{ISet} data sample.
    }} 
    \label{fig:StageC_MLP_response}
    }

\end{figure*}

\subsection{Additional SD-based variables}

With the goal of identifying muons with the scintillator layer of the ALTO array, we develop two simple variables based on SD information, and these are: the sum of all charges detected in the SDs for an event $Q_{\rm SD}$ and the number of SDs triggering an event $\log_{\rm 10}\left(N_{\rm SD}+10 \right)$. 

\begin{table}[t]
    \small
    \textbf{Definition of Bins in \bm{${\rm log}_{10}\left(N_{\rm pe}\right)$} and input characteristics}
    \begin{center}
        \begin{tabular}{c|cccc}
        \hline
        \textbf{Bin}  & \bm{${\rm log}_{10}\left(N_{\rm pe}\right)$}  & \textbf{Required} &   \bm{$N_{\rm train}$} & \bm{$N_{\rm test}$} \\ 
        
        \textbf{name} & \textbf{range} &  \textbf{signal \bm{$\varepsilon_{\rm Sig}$} (\%)} &  &\\
        
        \hline
        C1 & 1.00 -- 3.76 & 55 & 21421 & 14282 \\
        C2 & 3.76 -- 4.27 & 75 & 21321 & 14214 \\
        C3 & 4.27 -- 7.00 & 90 & 21235 & 14158 \\
        \hline

        \end{tabular}
    \end{center}

    \vspace{3mm}
    \textbf{Obtained and output characteristics}
    \begin{center}
        \begin{tabular}{c|rr|cc|rr}
        \hline
        \textbf{Bin} & \multicolumn{2}{c|}{\textbf{Obtained background}} & \multicolumn{2}{c|}{\textbf{Cut value }} & \multicolumn{2}{c}{\textbf{Selected events'}} \\ 
        \textbf{name} & \multicolumn{2}{c|}{\textbf{on test data \bm{$\varepsilon_{\rm BG}$} (\%)}} & \multicolumn{2}{c|}{\textbf{on test data}}&  \multicolumn{2}{c}{\bm{${\rm log}_{10}\left(E_{\rm T}\right)$} \textbf{mean \bm{$\pm$} sigma}}\\
        \hline
        & WCD+SD & WCD Only & WCD+SD & WCD Only & WCD+SD & WCD Only \\
        
        \hline
        C1 & $14.04\pm0.31$ & $15.63\pm0.33$ & 0.67 & 0.65 & $2.74 \pm 0.26$ & $2.75 \pm 0.26$\\
        C2 & $8.13\pm0.24$ & $8.82\pm0.25$ & 0.75 & 0.74 & $3.00 \pm 0.25$ & $3.00 \pm 0.25$\\
        C3 & $2.53\pm0.13$ & $3.28\pm0.15$ & 0.82 & 0.79 & $3.61 \pm 0.42$ & $3.61 \pm 0.42$\\
        \hline

        \end{tabular}
    \end{center}

    \caption{\small{\textbf{Stage C performance.} The first sub-table shows the definition of the bin ranges in $\log_{10}(N_{\rm pe})$ and the target $\gamma$-ray \emph{(signal)} efficiency, $\varepsilon_{\rm Sig}$, together with the number of $\gamma$ rays and protons \emph{(background)} in each bin which are input to the stage C test/train procedure. An equal number of events is used for \emph{signal} and \emph{background} training, given by $N_{\rm train}$, and similarly for $N_{\rm test}$. The second sub-table gives the results obtained at the end of stage C.  For each bin, this gives three pairs of columns showing the efficiency of \emph{background}, $\varepsilon_{\rm BG}$, the corresponding cut values, and the mean and sigma of the $\log_{10}(E_{\rm T} )$. The first/second columns in each pair show results with/without using the variables based on the scintillators in the train/test procedure. The statistics are obtained with the \emph{ISet} data sample.}}
    \label{table:GH_bins}
\end{table}

\subsection{Performance of stage C}
\label{sub_sec:stageC_performance}

The distribution of the fifteen variables used in the MLP train/test procedure are shown in figure~\ref{fig:StageC_variables} for the bin with the highest $N_{\rm pe}$,  C3. 
All the variables, except $\delta q_{\rm mean}$, are taken as $\log_{10}$ and re-scaled between the intervals [-1, 1] before the MLP train/test procedure.
Table~\ref{table:GH_bins} presents the performance of stage C, showing the decreasing  proton \emph{background} efficiency as a function of energy. 
The cut values are derived using a pre-defined signal efficiency and are based on the MLP response we obtain for each bin.
Figure~\ref{fig:StageC_MLP_response} shows the MLP classifier output, again for bin C3. 

In order to test the importance of the SDs in the $\gamma$-ray \emph{signal} over proton \emph{background} discrimination phase, a special MLP train/test is performed with the same configuration as described above, but without the two scintillator-based variables. The results are also presented in table~\ref{table:GH_bins}. 
In the lowest $N_{\rm pe}$ bin, C1, in the best-case of our analysis developments, our SD variables help only in rejecting an additional $\sim$1.5\% of proton \emph{background}.

\section{\textbf{Event energy evaluation in stage D}}
\label{sec:energy_reco}
The final stage of the analysis is to reconstruct the $\gamma$-ray energy using an MLP procedure in regression mode \cite{tmva}. 
In a supervised regression method, a list of predictive input variables is used to evaluate a continuous \emph{target} variable. 
The regression task assigns a target value from a list of provided input variables. 
The relevant information carried by the input variables needs to be correlated (or anti-correlated) with the target variable.
The 15 variables defined for stage C are also used as inputs in this regression method, while the target is the true energy of the $\gamma$ ray ($\log_{10}(E_{\rm T})$).
We keep the information carried by the SD variables also at this stage, because the total charge in the SDs ($Q\rm_{SD}$) and number of SDs triggered ($N\rm_{SD}$) are correlated with the energy of the event.
In this stage, we add a new variable helping in evaluating the footprint size of the shower in the array, see below, bringing the total number of variables used in this analysis step to 16.

\begin{figure}[t]
{\includegraphics[width=0.49\textwidth, height=0.45\textwidth]{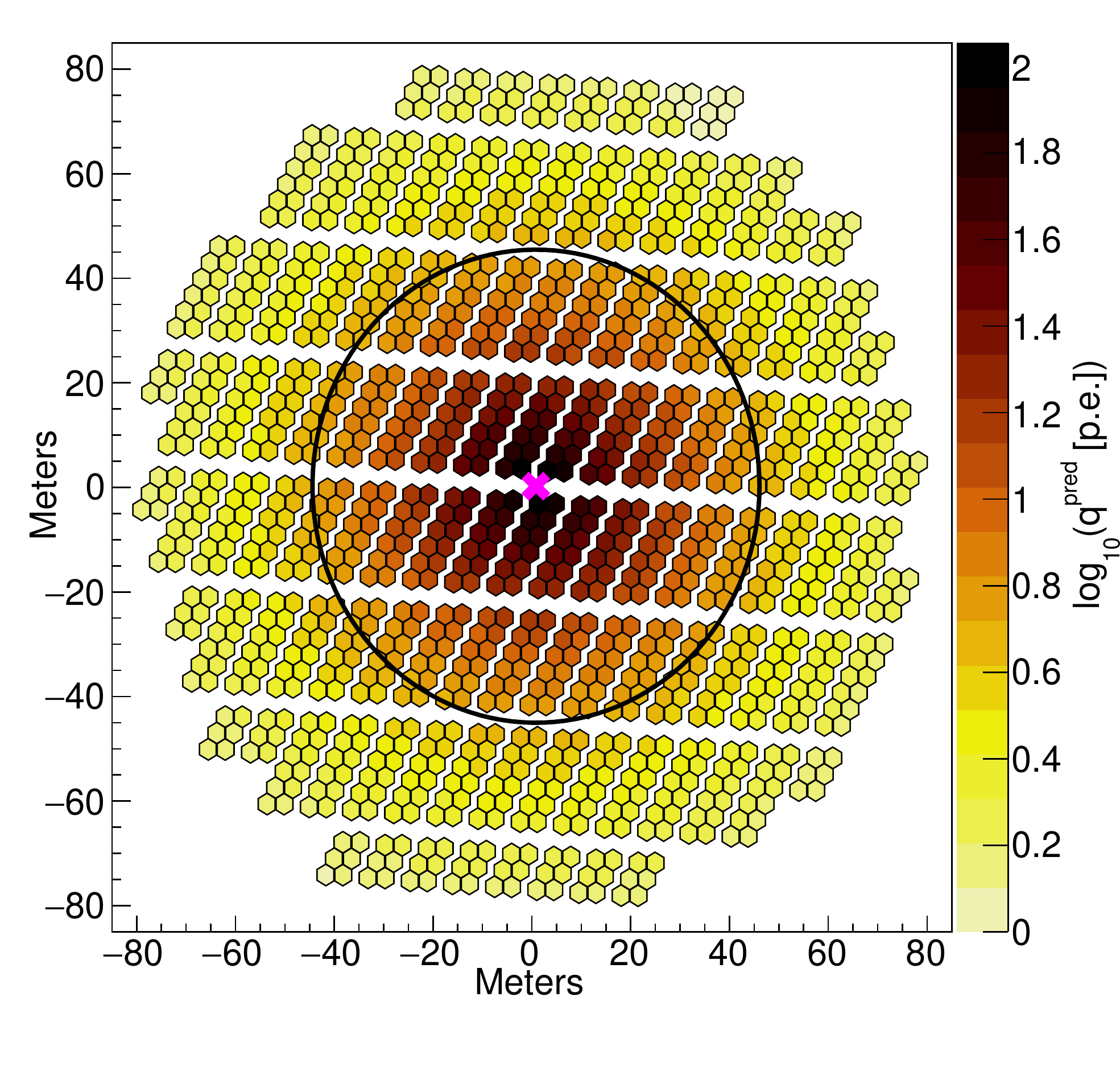}}
{\caption{\small{\textbf{Shower footprint size} \bm{$R_{\rm max}.$}
The predicted charge $q^{\rm pred}$ in all the WCDs is shown as viewed from the shower plane for an event having the reconstructed shower core at the centre of the array which is represented by a cross mark. The radius of the circle represents $R_{\rm max}$, which is the maximum distance between the reconstructed shower core and a WCD with $q_i^{\rm pred} > 5\,\rm{p.e.}$}}
\label{fig:max_exp_radius}}
\end{figure}

\subsection{Shower footprint size}
During the energy reconstruction procedure, we wish to access the information about the \emph{size} of the event as seen in the ALTO array and we do this by defining the \emph{shower footprint size}.
The predicted charges in all the ALTO detectors and the fitted parameters of the reconstructed shower, can be used to provide a predicted shower footprint size.
So, for each event and for each of the WCDs composing ALTO, we predict the charges $q^{\rm pred}_{i}$ to obtain a footprint of charges as shown in figure~\ref{fig:max_exp_radius}, where $i$ ranges from 1 to 1242. Using this information, the maximum predicted charge radius ($R_{\rm max}$) for an event is defined as the maximum distance from the reconstructed core where we find a WCD with $q^{\rm pred}_{i}>5\,\rm p.e.$ The \emph{predicted} shower footprint is simpler to calculate with respect to the \emph{detected} one, as the predicted charges can be calculated in each WCD composing ALTO.

\subsection{Performance of stage D}
\begin{figure}[t]
    \centering
    \includegraphics[width=0.48\textwidth]{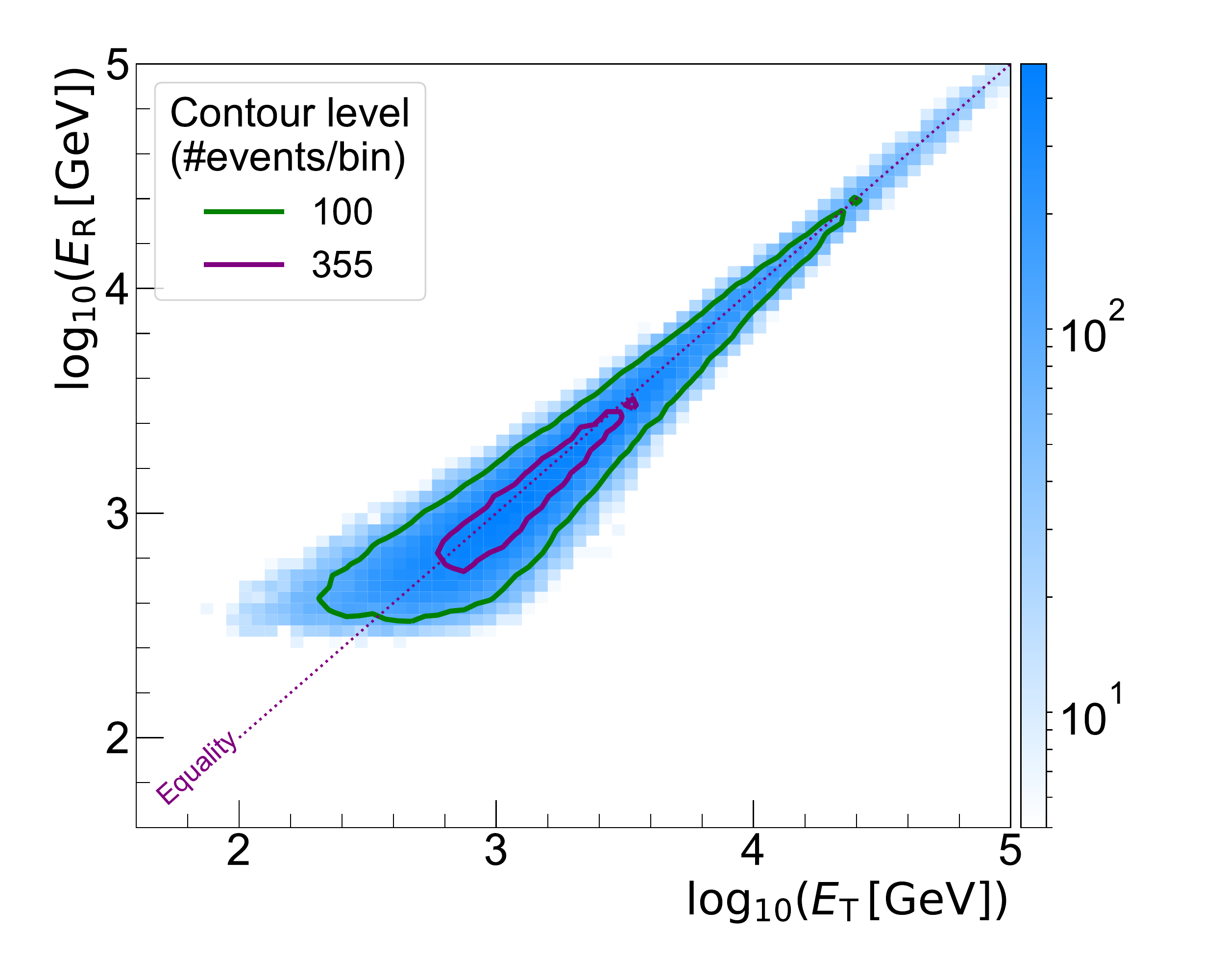}
    \hspace{0.1cm}
    \includegraphics[width=0.48\textwidth]{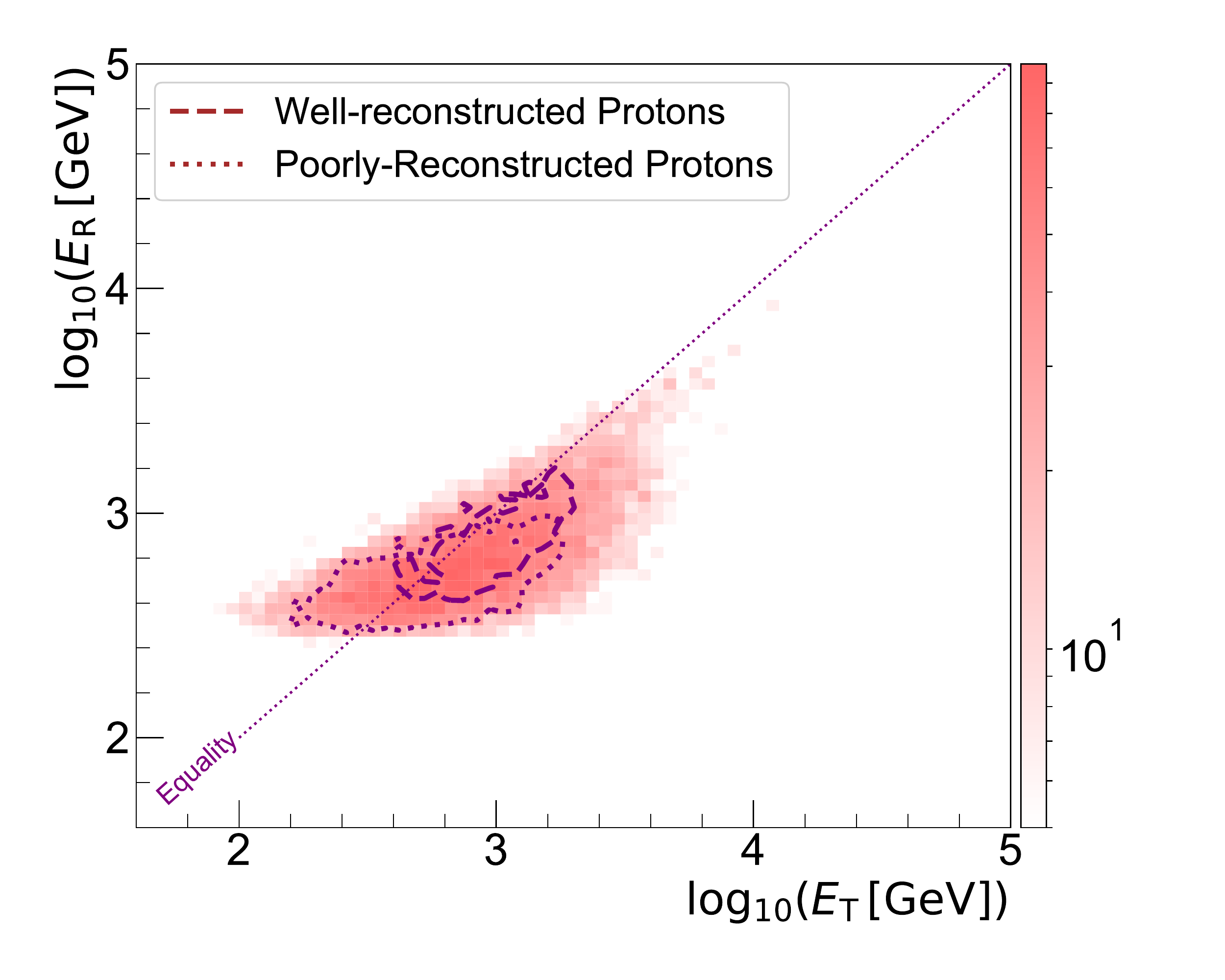}
    \caption{\small{\textbf{Stage D energy estimation for $\gamma$ rays and protons.} 
   \emph{Left panel:} The relation between the true energy, $\log_{10}(E_{\rm T})$, and the reconstructed energy, $\log_{10}(E_{\rm R})$, for $\gamma$-rays obtained in stage D. \emph{Right panel:} The same relation for the final 2\% fraction of remaining protons after the SEMLA analysis (colour scale), which contains 43\% of $\gamma$-like events (with dashed contours, representing the distribution of the ``Well-reconstructed'' subset)  and 57\% of poorly-reconstructed proton events (with dotted contours), with these contours at a 25-event/bin level. 
    The plots are obtained with the \emph{PSet} data sample.}
    }
    \label{fig:StageD_ERvsET}
\end{figure}

$\gamma$-ray events passing stage C are used as input in our regression model devoted to the energy evaluation of the event. 
We call the obtained MLP $\gamma$-ray energy as the reconstructed energy $E_{\rm R}$, see figure~\ref{fig:StageD_ERvsET} (left panel). 
This stage D energy-reconstruction MLP can be applied to evaluate the reconstructed energy also for proton events, \emph{right panel,} but it gives a biased evaluation of the energy in that case, because the regression model is obtained for $\gamma$ rays.   
So, with the stage D procedure we can assign a reconstructed energy to every event passing stage C, with no further selection of events for this stage.

\section{ALTO-array performance with SEMLA}
\label{sec:performance}

\begin{figure}[t]
    \centering
    \includegraphics[width=0.48\textwidth]{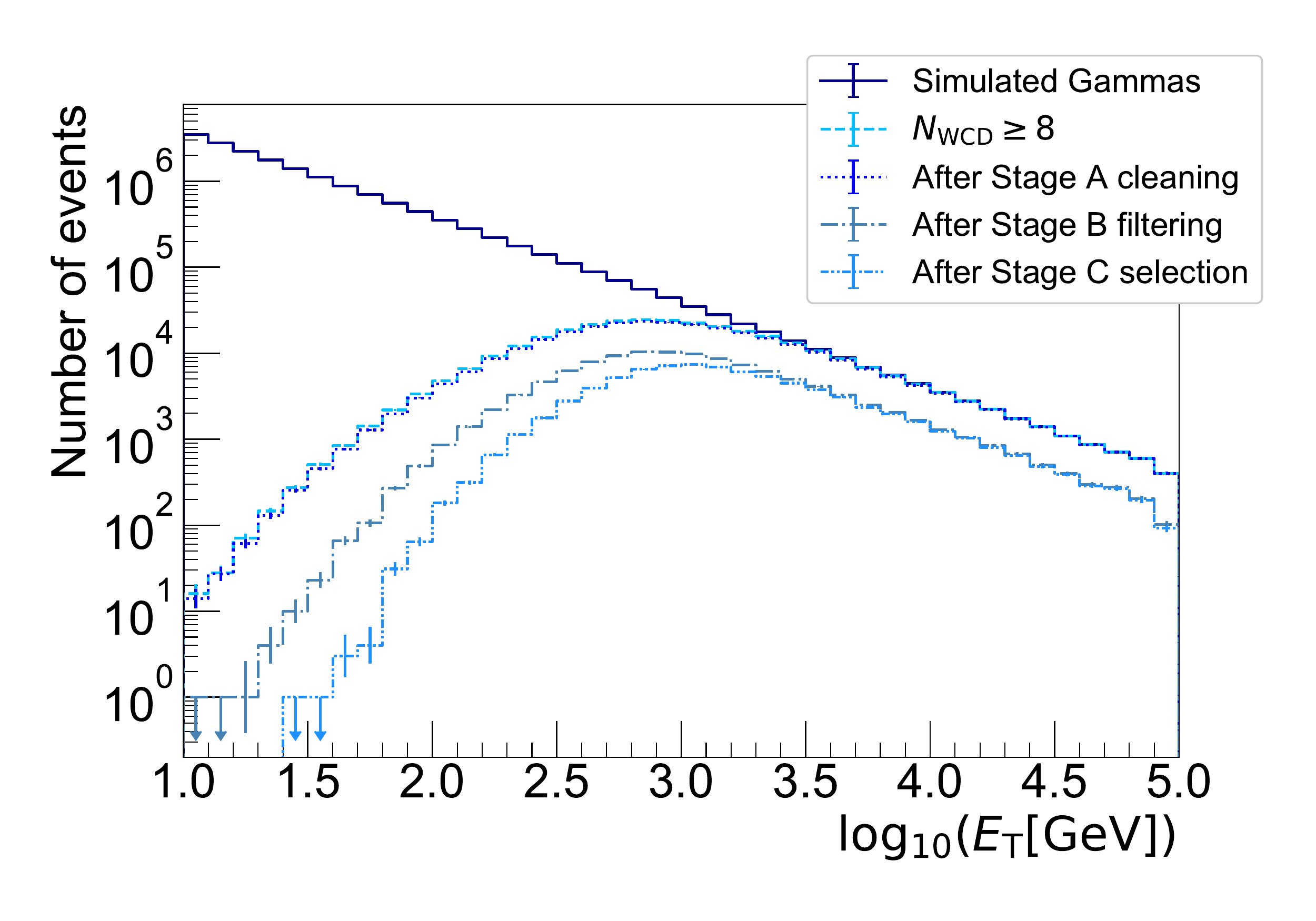}
     \hspace{.1
     cm}
     \includegraphics[width=0.48\textwidth]{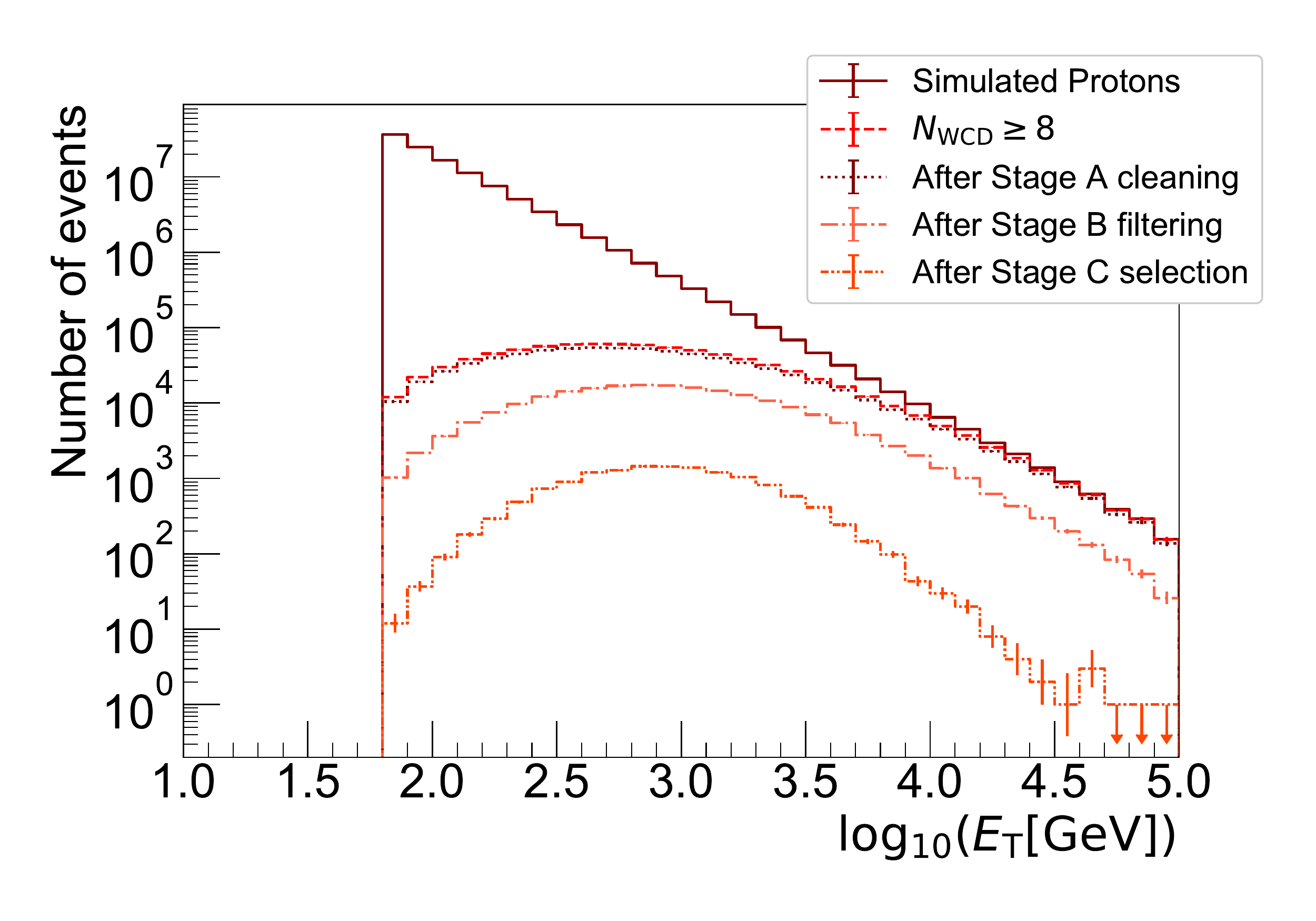}

    \caption{\small{\textbf{SEMLA performance on simulated $\gamma$ rays and protons.} 
    \emph{Left Panel:} True energy distribution of simulated $\gamma$-ray events after each SEMLA analysis stage.
    \emph{Right panel:} True energy distribution of simulated protons after each SEMLA stage. The plots are obtained with the \emph{PSet} data sample.}
    \label{fig:Energy_distributions}}
\end{figure}

We now evaluate the performance of the SEMLA analysis through an independent set of $\gamma$-ray and proton simulations, the \emph{PSet}. 
We apply the stage A method for an initial cleaning and then we subsequently apply stages B, C and D  using the obtained weights and analysis cuts. 
The effect of the consecutive analysis stages on the simulated $\gamma$-ray and proton event distributions are shown in figure~\ref{fig:Energy_distributions} left and right panels respectively, while table~\ref{table:event_stat} presents the fraction and the rate of remaining events after each stage.

In the case of $\gamma$-rays, taking as a reference the distribution of events passing stage A, the effect of stage B is to uniformly remove poorly-reconstructed events over the full energy range. 
Stage C cuts affect only the performance for $E\rm\lesssim1\,TeV$, since extracting $\gamma$-ray \emph{signal} from proton \emph{background} is harder in this region, while the higher-energy part instead remains unchanged. 
For protons, the effect of stage B is similar to that on $\gamma$-rays, while the effect of stage C is drastic, reducing the event rate in the full energy range by a factor of $\sim10$. After all analysis cuts, the cosmic-ray rate (scaled from the proton \emph{background} rate) is found to be $\sim\SI{181}{\Hz\per\steradian}$ (or $\SI{3.3}{{\min}^{-1} deg^{-2}}$) in the simulated zenith-angle range.
The $\gamma$-ray \emph{signal} and the proton \emph{background} distributions both have a maximum at $\sim\SI{1}{\TeV}$.

Applying also to the proton events the three conditions for the definition of a well-reconstructed event given in section\ \ref{sub_sec:precut_input_def}, we can now study the contamination of poorly reconstructed events in the final samples.
Of the remaining 25\% of $\gamma$-rays after all analysis cuts, 95\% are well-reconstructed events, while the remaining 5\% are poorly-reconstructed events. 
For protons, the final 2\% fraction of events, see figure~\ref{fig:StageD_ERvsET} (right panel), is composed by an intrinsically irreducible 43\% of $\gamma$-like events, while the remaining 57\% of events could be possibly suppressed by adding more information, for example provided by more sophisticated detector units.
This demonstrates that with SEMLA we have been able to achieve an almost optimal detection of well-reconstructed $\gamma$-rays. 

\begin{figure}
{\includegraphics[width=0.56\textwidth]{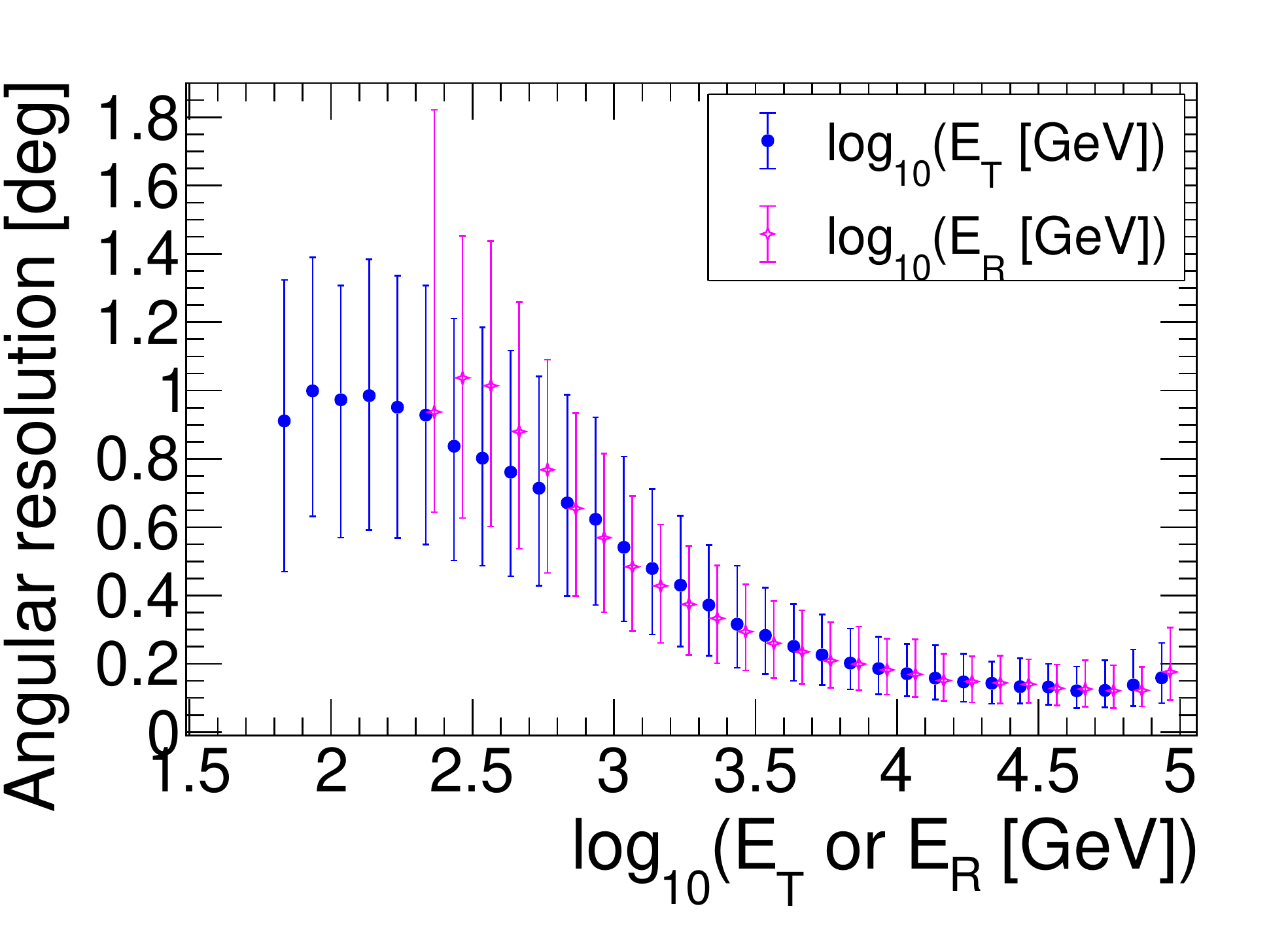}}
{\caption{\small{\textbf{Angular resolution of simulated $\gamma$-rays selected by SEMLA.} 
    The angular resolution in degrees, see section\ \ref{sec:performance}.  For clarity, the points in $E_{\rm T}$ and $E_{\rm R}$ are slightly shifted left and right respectively in energy. The plot is obtained with the \emph{PSet} data sample.
    }}
    \label{fig:StageC_selected_angular_resolution}}
\end{figure}

Figure~\ref{fig:StageC_selected_angular_resolution} shows the angular resolution (in degrees) for simulated $\gamma$-rays after the SEMLA analysis. Here, the angular resolution represented is the 68.3\% containment of the distribution of the angle between each event's reconstructed shower direction and the true (point-source) direction.
The error bars are indicative of the spread, showing the 38.3\% to 86.6\% containment range.
The performance of ALTO in terms of angular resolution improves from $\sim0.8^\circ$ at $300\rm\,GeV$ to $\sim 0.15^\circ$ at $10\rm\,TeV$.
This 68.3\% containment as a function of reconstructed energy $E_{\rm R}$ defines an angular cut on the events depending on which bin in $E_{\rm R}$ contains them. This selects an optimal angular acceptance region for a point-like source while minimizing the contamination from the diffuse cosmic-ray background, as shown in \cite{KlepserPSF}.

\begin{table}[t]
    \small
    \centering
    \begin{tabular}{c|rr|cc}
        \hline

        \textbf{Stages} & \textbf{$\gamma$-ray}  & \textbf{Proton} & \textbf{$\gamma$-ray rate} & \textbf{Cosmic-ray rate} \\
        & \textbf{fraction}  & \textbf{fraction} & $\mathbf{[min^{-1}]}$ & $\mathbf{[deg^{-2}min^{-1}]}$\\
        \hline
            Stage A & 95.2\% & 88.9\%  & 2.11 & 171.0 \tabularnewline 
            Stage B & 36.6\% & 25.8\% & 0.81 & 49.2 \tabularnewline 
            Stages C--D & 25.1\% & 1.7\% & 0.56 & 3.3 \tabularnewline 
        \hline
    \end{tabular}
    
    \caption{\small{\textbf{SEMLA efficiencies and indicative rates.} Remaining fraction of simulated $\gamma$-rays and protons relative to the events with $N_{\rm WCD}$ $\geq$ 8 after each SEMLA stage, before the application of the angular selection cut.  The $\gamma$-ray rates are estimated for the Pseudo-Crab source, while the cosmic ray rates result from rescaling the protons to the cosmic-ray flux, as described in section \ref{subsec:sim_atmos}. The values are obtained with the \emph{PSet} data sample.
    }}
    \label{table:event_stat}
    
\end{table}

\begin{figure}[t] 
    \centering
    \includegraphics[width=0.475\textwidth]{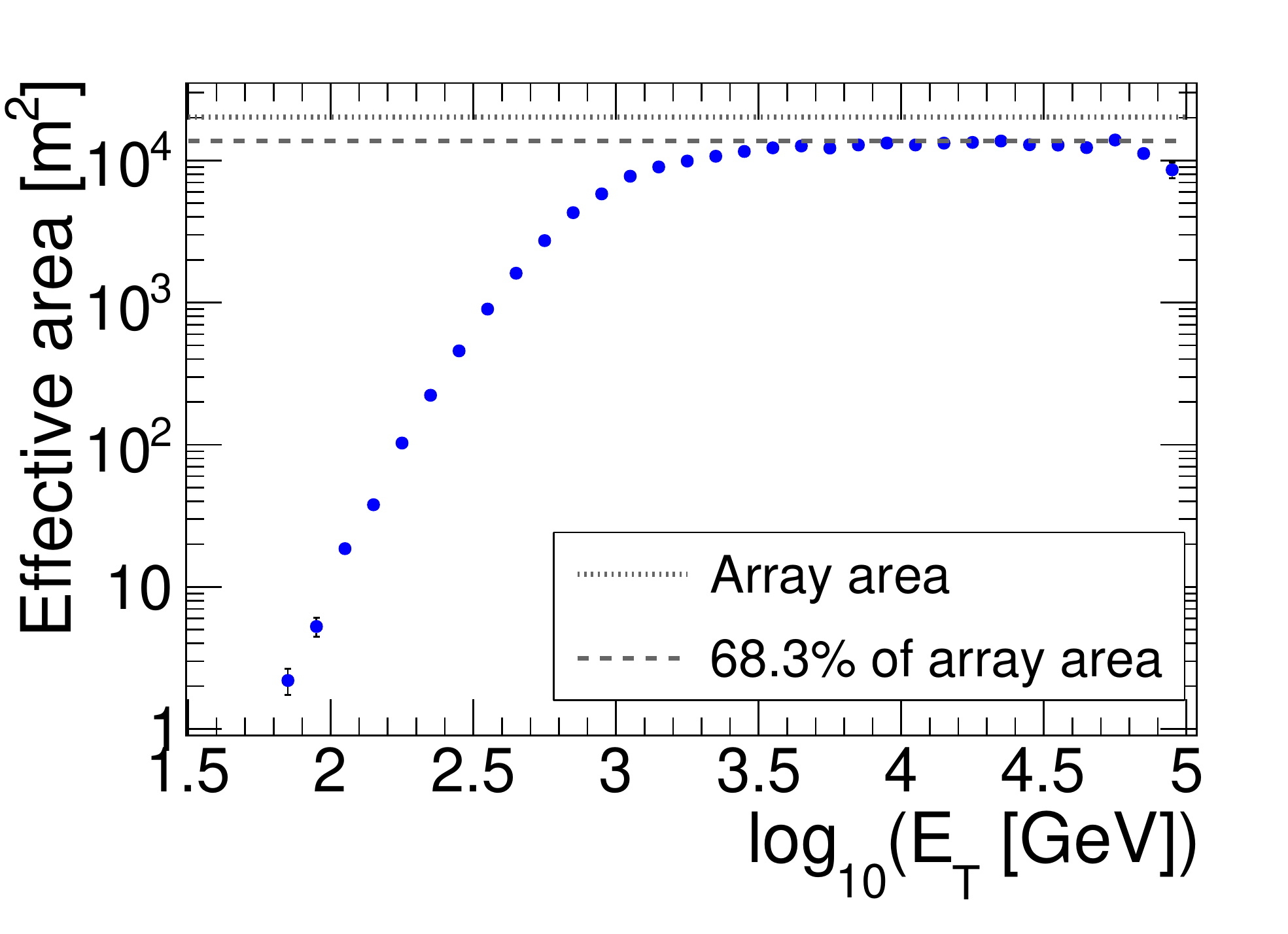}
    \hspace{0.5cm}
    \includegraphics[width=0.475\textwidth]{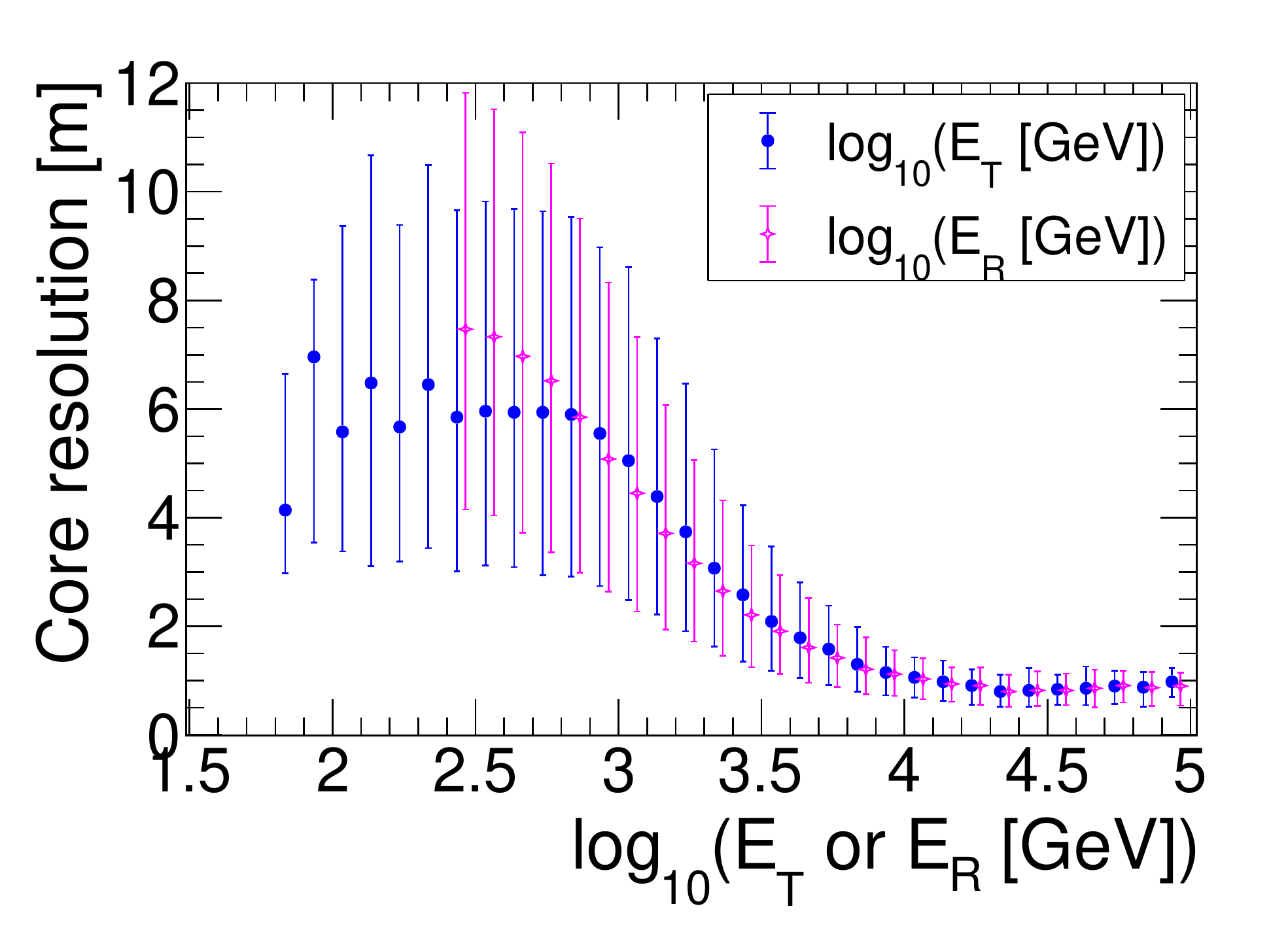}
    
    \includegraphics[width=0.475\textwidth]{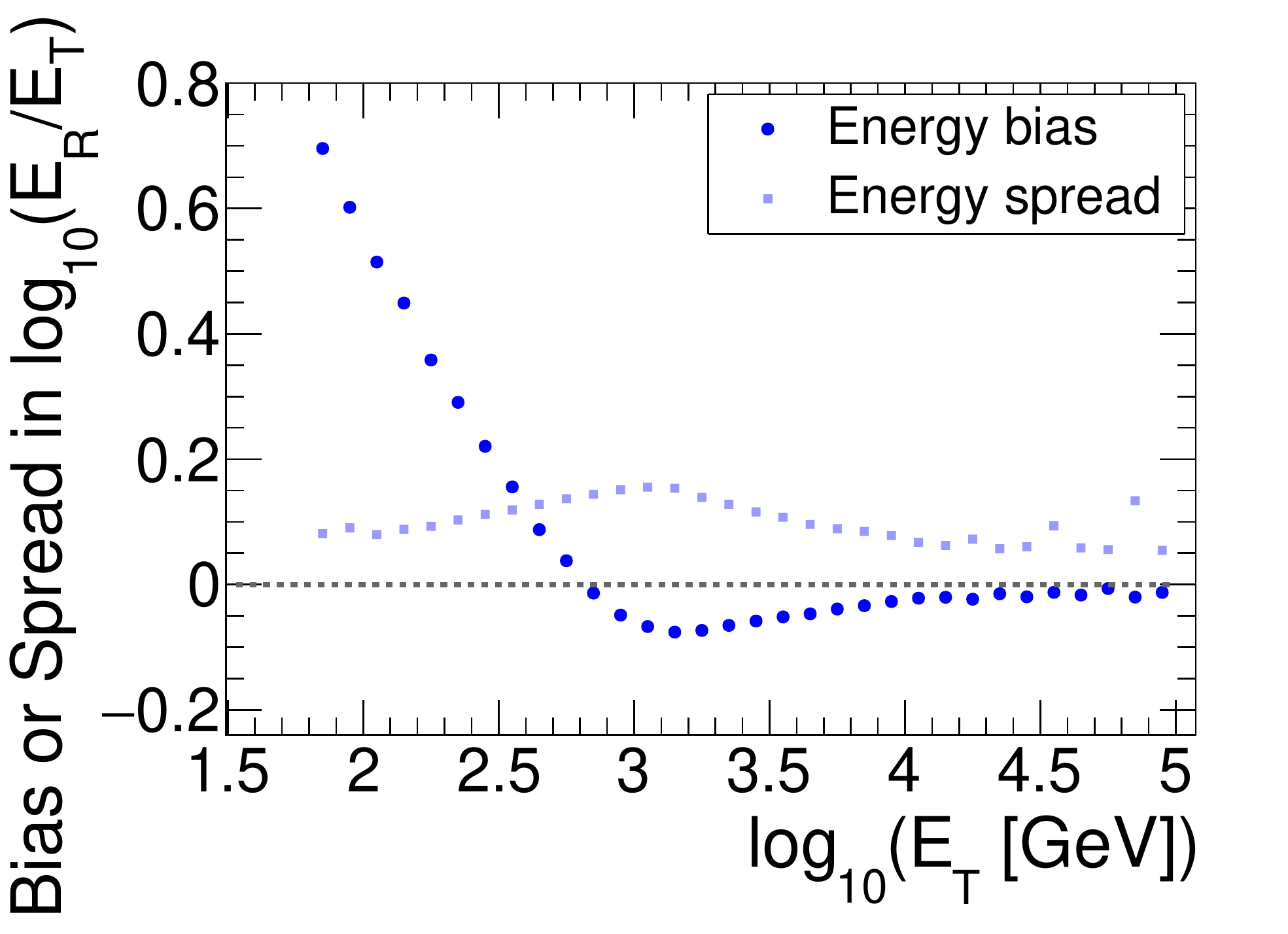}
    \hspace{0.5cm}
    \includegraphics[width=0.475\textwidth]{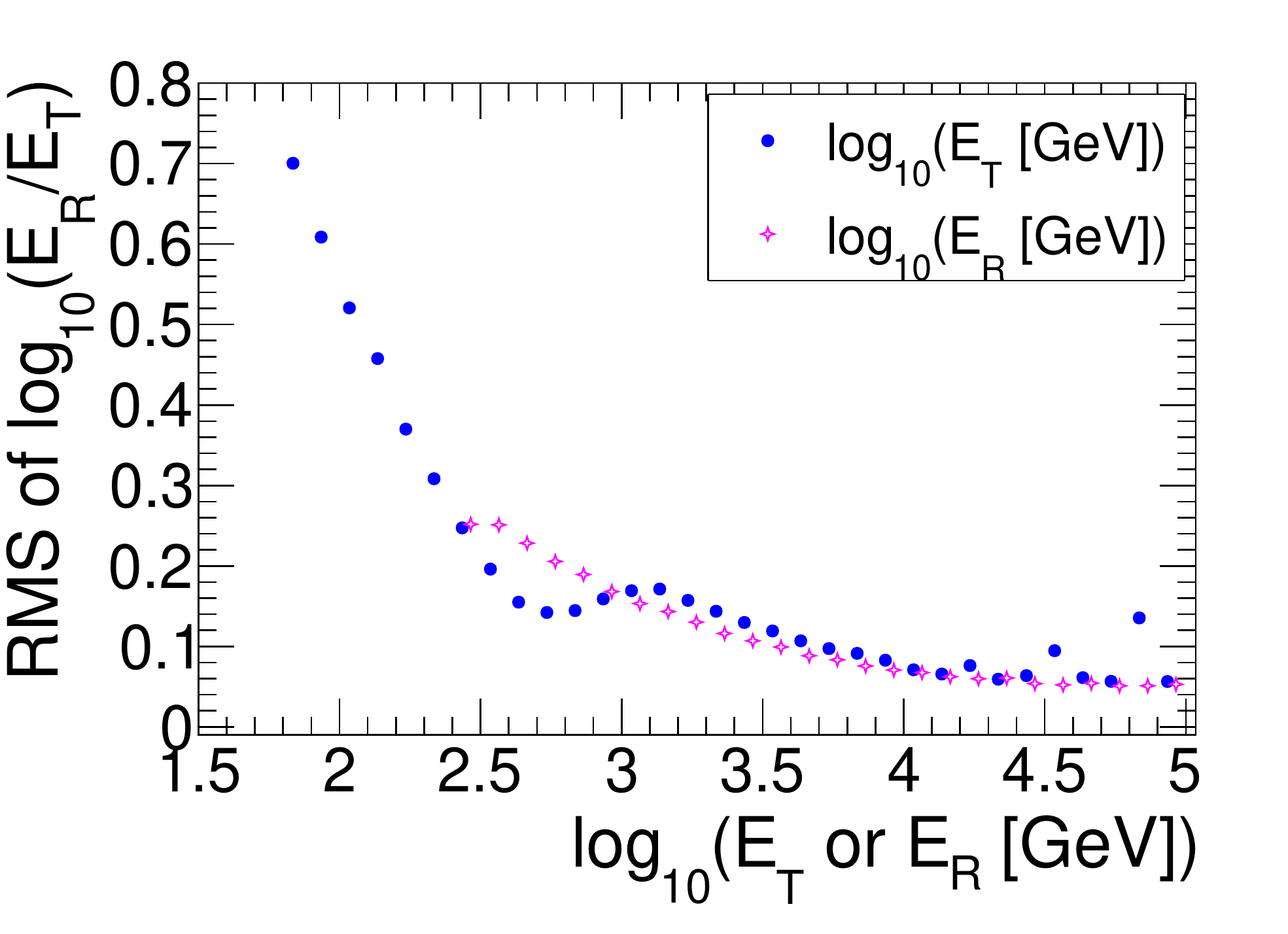}
    
    \caption{\small{\textbf{SEMLA performance on simulated $\gamma$-rays after angular selection cut.}
    \emph{Upper Panel:} The effective area in $\rm m^2$ (left) and the core resolution in metres (right).
    \emph{Lower Panel:} The relative energy bias $\overline{\Delta_{\small 10}E}$ and its spread $\sigma_{\Delta_{\small 10}E}$ (left), and the energy resolution (the RMS of $\Delta_{\small 10}E$) (right), as described in section~\ref{sec:performance}. In the right panels, the points in $E_{\rm T}$ and $E_{\rm R}$ are slightly shifted left and right respectively in energy for clarity. The plots are obtained with the \emph{PSet} data sample.
    }}
    \label{fig:semla_perf_plots_angcut}
\end{figure}

The final $\gamma$-ray performance for events passing all SEMLA stages, and in addition passing this angular selection cut given by the crosses in figure~\ref{fig:StageC_selected_angular_resolution}, is shown in figure~\ref{fig:semla_perf_plots_angcut} in terms of effective area, energy and core resolution, and energy bias, spread, and resolution.  For the $\gamma$ rays, the angular selection cut simply reduces their number by 31.7\% over the whole reconstructed energy range.
We note that the performance concerning the energy resolution and bias is practically unaffected by this angular selection cut --- since the angular and energy reconstructions are quite decoupled --- whereas the core resolution is somewhat improved.

Figure~\ref{fig:semla_perf_plots_angcut} (upper left panel) shows the final effective area for simulated $\gamma$-rays. The effective area is defined as the area of the simulation (in $\rm m^2$) multiplied by the ratio of the number of events passing the SEMLA analysis and the angular selection cut to the total number of events simulated.
The smaller effective area at low energies is due to a small number of triggered detectors and to the \emph{background} rejection procedure, while the plateau reached above $1\rm\,TeV$ is due to the limited size of the ALTO array (this area being shown by the dotted horizontal line, while the dashed line shows the decrease by 31.7\% due to the angular selection cut). 

Figure~\ref{fig:semla_perf_plots_angcut} (upper right panel) then shows the core resolution (in m) for simulated $\gamma$-rays.  Similarly to the angular resolution, the core resolution is shown as the 68.3\% containment distance between the true and reconstructed core positions, with the error bars indicating the spread between the 38.3\% and 86.6\% containment values.

As in \cite{HAWC_Crab_2019}, we define the energy bias and spread as the mean and standard deviation of the distribution of $\Delta_{\small 10}E = \log_{10}\left( \sfrac{E_{\rm R}}{E_{\rm T}} \right)$ in $\log_{10}(E_{\rm T})$ bins. 
The energy resolution is taken as the RMS of $\Delta_{\small 10}E$ in bins in $\log_{10}(E_{\rm T}\,{\rm\underline{or}}\,E_{\rm R})$. 
The mean, standard deviation, and RMS are given by: 

\begin{equation}
    \overline{\Delta_{\small 10}E} = \frac{1}{N} \sum_{i=1}^{N}\Delta_{\small 10}E_i \quad
\end{equation}

\begin{equation}
    \sigma_{\Delta_{\small 10}E} = \sqrt{ \frac{1}{N} \sum_{i=1}^{N}({\Delta_{\small 10}E}_i - \overline{\Delta_{\small 10}E})^2}
\end{equation}

\begin{equation}
    {\rm RMS}_{\Delta_{\small 10}E} = \sqrt{ \frac{1}{N}\sum_{i=1}^{N}(\Delta_{\small 10}E_i)^2}
\end{equation}
where $N$ is the number of events in the bin.  These are shown in figure~\ref{fig:semla_perf_plots_angcut} (lower panels).

In \cite{cta_energy_res} they define the energy resolution instead as the 68.3\% containment for $(E_{\rm R}-E_{\rm T})/E_{\rm T}$ in the half interval around 0.  This definition is in linear space rather than the $\log_{10}$ space above.\footnote{Conversion from $\log_{10}$ space to linear space is simply --- in the first approximation --- multiplication by $\ln(e) \simeq 2.3$, within 10\% precision for values $<0.2$, which is the case for the energy range above $300\,\rm GeV$.}  We note that, in our case, the post-SEMLA-cut distributions are quite Gaussian, so the RMS and 68.3\% are very close.

When $E_{\rm T} < 1\,\rm TeV$, only those showers 
with upward fluctuations in their number of particles 
are detected, due to the effect of the detection threshold.
So, the range of reconstructed energies is in fact restricted, with the resulting reconstructed energies always above $\sim250\,\rm GeV$, see figure~\ref{fig:StageD_ERvsET}.
This threshold effect leads to a decrease in the error on the energy for $E_{\rm T} < 1\,\rm TeV$.

The final $\gamma$-ray rate from the Pseudo-Crab source after all analysis cuts, including the angular selection cut for the point-like source, is $\sim 0.38\; \rm min^{-1}$ ($\sim 23$ per hour), while the expected cosmic-ray background within the angular selection cut (scaled from the proton background rate) is $\sim 5.6\; \rm min^{-1}$ ($\sim 336$ per hour).

\section{Summary and outlook}
\label{Outlook}

In this paper, we have presented the SEMLA data analysis method for the detection of atmospheric $\gamma$-ray showers with the ALTO detector. We underline that the analysis technique we propose can be easily applied to any other type of experiment, provided that the experiment-relevant variables are implemented and used.

The analysis method contains four stages, each devoted to a particular task. 
The first stage simply removes events containing abnormal values of the reconstructed variables (stage A).
The two following stages make use of machine learning for the classification of well- versus poorly-reconstructed $\gamma$-ray events (stage B) and for the classification of $\gamma$-rays versus protons (stage C). 
The final stage consists in the evaluation of the energy of the events, considering the events as $\gamma$-rays. 
Therefore, the energy evaluated for protons is obtained considering them as being $\gamma$-rays. 

The final performance of the analysis can be seen in figures~\ref{fig:StageD_ERvsET}, \ref{fig:Energy_distributions}, \ref{fig:StageC_selected_angular_resolution} and \ref{fig:semla_perf_plots_angcut}. 
Overall, we obtain a good performance both in \emph{background} suppression and energy reconstruction, given the chosen shower reconstruction procedure.

However, the performance for $E<800 \,\rm GeV$ could be further improved with respect to what we present here in this paper. 
An important consideration for the optimisation of the lowest energies is the triggering strategy. In this paper we show the performance of a ``2/12'' trigger, meaning that two detectors out of 12 composing an ALTO cluster must have a signal amplitude above $20\,\rm mV$. 
The chosen trigger configuration favours the detection of muons present in the proton showers, as downward-going muons will produce a detectable signal in one WCD and the SD beneath it.
By changing the trigger conditions, as for instance using a ``2/6'' or a ``1/6'' condition based only on WCDs in an ALTO cluster, the low-energy performance efficiency can be improved, with the drawback that the global event rate will increase. 
Dependent on the capability of the data-acquisition system to handle higher trigger and data rates, the sensitivity to lower energies could therefore be further optimised.

In section \ref{sec:performance} we also find that about 43\% of the remaining proton sample is still composed by ``clean'' events having good angular and 
core resolutions, and a good energy reconstruction, 
while about 57\% of the protons could in principle still be suppressed as they instead have bad angular, core and energy resolutions. 
This 57\% of poorly-reconstructed protons are in the low-energy region where differences in the time and charge footprints of $\gamma$-rays and protons can hardly be seen, but 
there are more advanced approaches which could help reaching an improved event classification in this energy region. 
A first possibility is on the hardware side, by increasing the performance of the detector units, for instance by choosing a PMT with better timing resolution or by adding more PMTs in a single unit. 
Having a better PMT could help in achieving an enhanced reconstruction accuracy, and adding more PMTs could help in the detection of fainter signals.
A second possibility is a finer granularity of the ALTO array which might help obtaining a more detailed shower footprint by avoiding big gaps in the images. 
And finally, a third possibility could concern the reconstruction and the analysis themselves, by making use of Deep Learning methods for regression and classification \cite{dnn}, 
or by using template-based global likelihood reconstruction methods \cite{Mpp, ImPACT}. 
All these possible improvements might increase the performance but at the price of a higher cost for the project.

With a fully developed analysis chain we are able to investigate the importance of several design parameters associated with the ALTO detector, discussion of which is beyond the scope of this paper. As an example, we demonstrated in section \ref{sub_sec:stageC_performance} that for the purpose of $\gamma$-ray astronomy, there is no significant additional \emph{background} discrimination power from secondary muon detectors implemented as scintillators positioned underneath the WCDs and separated by $25\rm\,cm$ of concrete. Therefore, an ALTO unit composed by only a primary detector (either a WCD alone or an improved version of an SD alone) will keep the project costs down and will simplify the operation of the array. 

The differential sensitivity of the ALTO array with different trigger conditions and reconstruction settings, together with the expected detection performance on a list of astrophysical $\gamma$-ray point-like sources, and the discussion on the ALTO design, including the relevance of secondary muon detectors, are the subject of a forthcoming paper where the SEMLA analysis scheme, presented here, will be used.

\section*{\ack{}}

The ALTO R\&D project is being primarily supported by Linnaeus University and the following Swedish private foundations or public institutes: the Crafoord Foundation, the Foundation in memory of Lars Hierta, the Magnus
Bergvall’s Foundation, the Crafoord stipendium of the Royal Swedish Academy of Sciences (KVA),
the Märta and Erik Holmberg Endowment of the Royal Physiographic Society in Lund, the Längmanska kulturfonden, the Helge Ax:son Johnson’s Foundation and Linnaeus University. We also
thank the Swedish National Infrastructure for Computing (SNIC) at Lunarc (Lund, Sweden). 
The authors would like to thank the \emph{Data Intensive Sciences and Applications} (DISA) research centre at Linnaeus University for their valuable support.
We also acknowledge funding from the Abu Dhabi Award for Research Excellence (AARE-2019) under the project number AARE19-224.
The authors would like to thank Staffan Carius for reading the paper and providing constructive feedback.
We thank GNU parallel \cite{parallel} which enabled us to run the simulation in parallel in our local cluster.  
We wish to thank the anonymous referee and editor for the helpful comments which clarified some points in the paper.

\providecommand{\planckit}{\textit}

\end{document}